\documentclass[twocolumn,amsmath,amssymb,showpacs,nofootinbib]{revtex4-1}

\usepackage[usenames,dvips]{color}
\usepackage{graphicx}
\usepackage{dcolumn}
\usepackage{bm}
\usepackage{epstopdf}
\usepackage{xcolor}
\usepackage{dsfont}

\newcommand{\bea}{\begin{eqnarray}}
\newcommand{\eea}{\end{eqnarray}}

\newcommand{\phd}{\phantom{\dag}}
\newcommand{\ph}{\phantom{.}}

\newcommand{\noi}{\noindent}
\newcommand{\no}{\nonumber}
\renewcommand{\Re}{\operatorname{Re}}
\renewcommand{\Im}{\operatorname{Im}}

\begin{document}

\title{Majorana fermion fingerprints in spin-polarised scanning tunneling microscopy}
\author{Panagiotis Kotetes$^{1}$}
\email{panagiotis.kotetes@kit.edu}
\author{Daniel Mendler$^{1,2}$}
\author{Andreas Heimes$^{1}$}
\author{Gerd Sch\"{o}n$^{1,2}$}
\affiliation{$^1$Institut f\"{u}r Theoretische Festk\"{o}rperphysik, Karlsruhe Institute of Technology, 76131 Karlsruhe, Germany}
\affiliation{$^2$Institute of Nanotechnology, Karlsruhe Institute of Technology, 76344 Eggenstein-Leopoldshafen, Germany}

\vskip 1cm
\begin{abstract}
We calculate the spatially resolved tunneling conductance of topological superconductors (TSCs) based on ferromagnetic chains, measured by means of spin-polarised scanning
tunneling microscopy (SPSTM). Our analysis reveals novel signatures of MFs arising from the interplay of their \textit{strongly anisotropic} spin-polarisation and the
magnetisation content of the tip. We focus on the deep Yu-Shiba-Rusinov (YSR) limit where only YSR bound states localised in the vicinity of the adatoms govern the low-energy
as also the topological properties of the system. Under these conditions, we investigate the occurence of zero/finite bias peaks (ZBPs/FBPs) for a single or two coupled TSC
chains forming a Josephson junction. Each TSC can host up to two Majorana fermions (MFs) per edge if chiral symmetry is preserved. Here we retrieve the conductance for all the
accessible configurations of the MF number of each chain. Our results illustrate innovative spin-polarisation-sensitive experimental routes for arresting the MFs by either
restoring or splitting the ZBP in a predictable fashion via: i) weakly breaking chiral symmetry, e.g. by the SPSTM tip itself or by an external Zeeman field and ii) tuning the
superconducting phase difference of the TSCs, which is encoded in the 4$\pi$-Josephson coupling of neighbouring MFs.

\end{abstract}


\maketitle

\section{Introduction}

Recent experiments in hybrid superconducting devices \cite{NWExp,Yazdani_Science,Basel} have revealed strong evidence for the existence of Majorana fermions (MFs)
\cite{KitaevUnpaired,TSCs,Kotetes}, which are neutral non-abelian zero-energy quasiparticle excitations in many-particle systems. The unambiguous confirmation of their
discovery and their successful experimental manipulation will be an evidence for this exotic kind of statistics and in addition will constitute fertile ground for
de\-ve\-lo\-ping topological quantum computing \cite{TQC}. The first experiments claiming the discovery of MFs have been performed in platforms based on semiconducting
nanowires \cite{NWExp} with strong spin-orbit coupling (SOC), a type of artificial topological superconductors (TSCs) predicted in Ref.~\onlinecite{NW}. One of the most
distin\-ctive features of an isolated MF is the appearance of a zero bias peak (ZBP) of height $2e^2/h$ in the tunneling conductance \cite{PALee,FlensbergTunneling}. However,
a ZBP is not unique to MFs \cite{ZBPAlternatives} and thus a number of experiments \cite{NWExpDebate} have scrutinised the findings concerning the possible presence of MFs in
such devices. 

\begin{figure}[t]
\centering
\includegraphics[width=1\columnwidth]{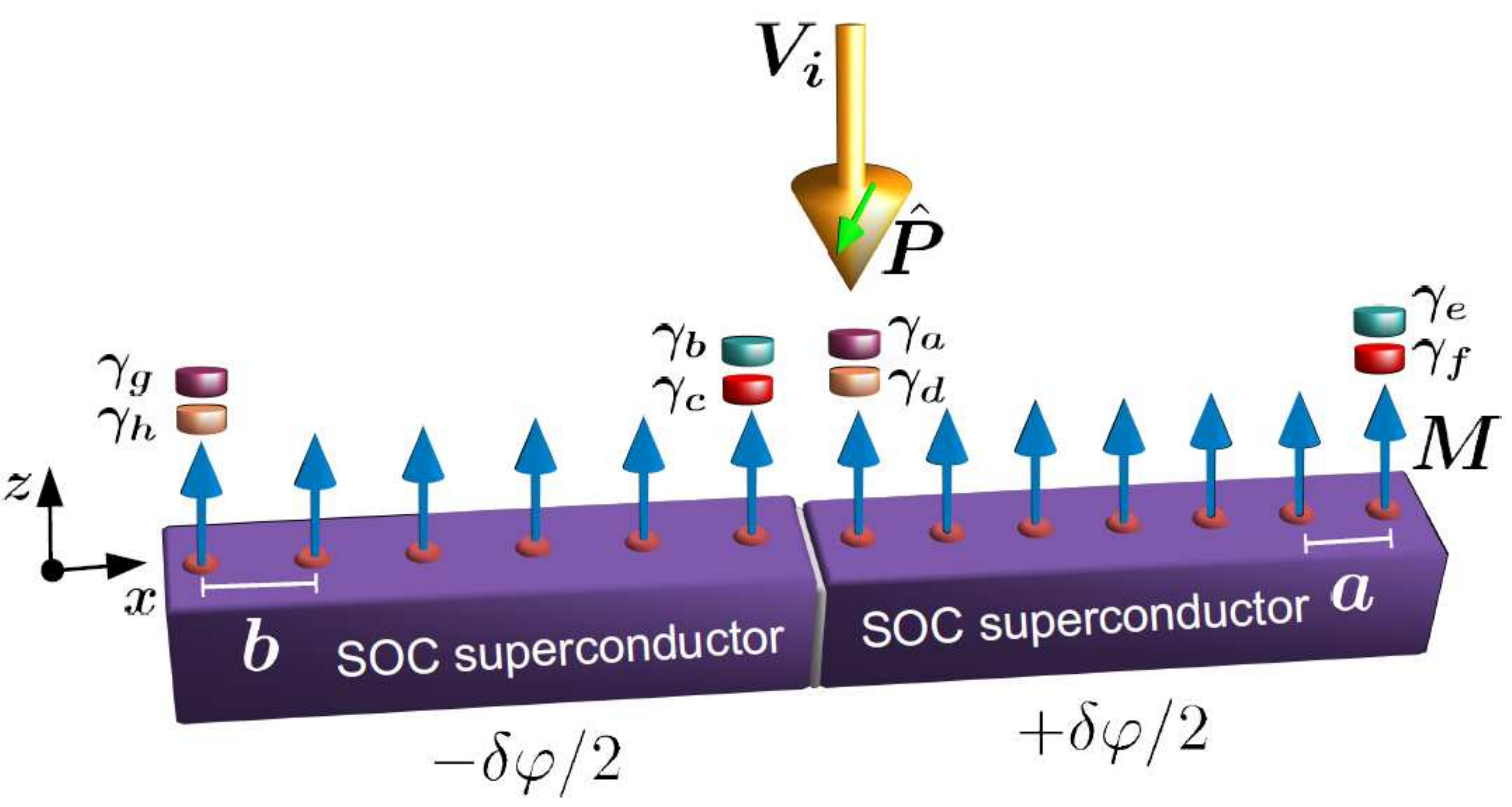}
\vspace{0.1in}
\caption{Two coupled topological superconductors (TSCs) based on magnetic chains probed by means of spin-polarised scanning tunneling microscopy (SPSTM). The tip electrons
owe a spin-polarisation along the $\hat{\bm{P}}$ direction. Each chain can host 1 or 2 MFs per edge by virtue of chiral symmetry. The two TSCs are interfaced by a thin
insulator yielding a Josephson junction since the superconducting phases on the two chains differ by $\delta\varphi$. By tuning the orientation of $\hat{\bm{P}}$, the value of
$\delta\varphi$, as also by controllably violating or restoring chiral symmetry, one can imprint Majorana fermion (MF) signatures on the tunneling conductance $dI_i/dV_i$.}
\label{fig:dIdV3d}
\end{figure}

The particular debate opened the door to new ideas for engineering MFs without involving semiconductors but by instead employing devices based on conventional SCs in the
presence of some kind of magnetic texture. One of the first proposals along this direction \cite{Choy} pinpointed that a magnetic chain with randomly ordered spins deposited
onto a SC can harbor MFs at its edges. Later on it was realised \cite{Martin,Kjaergaard2012} that in the absence of SOC and other external fields the presence of a helical
magnetic texture is the minimal requirement for engineering a TSC \cite{Kotetes}. More importantly, TSCs from magnetic chains on superconducting substrates can be manipulated
and probed using spin-polarised scanning tunneling microscopy (SPSTM) \cite{Yazdani}, allowing the spatial visualisation of the MF wavefunctions and thus providing a more
reliable method for diagnosing TSCs. This advantage further motivated a plethora of theoretical pro\-po\-sals involving helical magnetism \cite{Helical}. Remarkably, it has
been also shown that helicity (or SOC) can be artificially engineered using suitable external fields for the case of antiferromagnetically ordered chains \cite{Heimes}. If
instead SOC is present, both ferromagnetic \cite{Brydon,MacDonald,Interplay,Tanaka2014,Peng,ZBPDSarma,LutchynMultiChannel,Cadez} and antiferromagnetic
\cite{Interplay,Tanaka2014} chains on a SC can exhibit topologically non-trivial properties and host MFs. As a matter of fact, the former situation appears to have been
recently realised in the laboratory \cite{Yazdani_Science}. Nonetheless, further experiments at lower tem\-pe\-ra\-tu\-res and with improved energy resolution are required
for confirming the presence of MFs \cite{Yazdani_Science,ZBPDSarma}. 

In this work we illustrate a number of yet unexplored experimental fingerprints of MFs, part of which can be directly tested in the existing devices \cite{Yazdani_Science},
thus promising an unambiguous identification of the MFs. Key feature of our approach is properly taking into account the magnetic spin-polarisation content of the SPSTM tip,
which could also provide an alternative explanation for the very low value of the measured conductance in the above mentioned experiment. Previous studies in such systems have
focused on a non-magnetic tip in the normal \cite{Peng,ZBPDSarma} or superconducting phase \cite{SCtip}. Note however that spin-selective Andreev processes due to MFs have
been previously studied in Ref.~\onlinecite{SESARS} for nanowire-based TSCs. 

Here, following the method of Ref.~\onlinecite{FlensbergTunneling} we retrieve the tunneling conductance for ferromagnetically ordered chains within the microscopic
Yu-Shiba-Rusinov (YSR) \cite{Yu,Shiba,Rusinov,YazdaniShiba} type of model, extracted in Ref.~\onlinecite{Interplay}. The latter electronic bound states dominate the low-energy
properties of the substrate SC in the presence of the magnetic adatoms. If the atomic spins can be treated as classical, the YSR states become fully responsible for the
topological pro\-per\-ties of the hybrid device. As it has been shown in Ref.~\onlinecite{Interplay} such topological YSR bands can harbor 1 or 2 MFs per edge in the presence
of chiral symmetry. 

Specifically, we investigate the tunneling conductance properties and the emerging zero/finite bias peaks (ZBPs/FBPs) for i) isolated chains supporting either number of MFs
per edge and ii) pairs of coupled chains with the same or different number of MFs per edge. Central feature of our analysis is the electronic spin-po\-la\-ri\-sa\-tion induced
by the MFs, which exhibits a strongly anisotropic coupling to the tip electrons, transferring this distinctive behaviour to the tunneling conductance. Evenmore, depending on
the spin-polarisation of the magnetic tip, chiral symmetry can be weakly violated locally in a controlled manner. The latter can drastically affect the conductance spectra
and the qualitative features of the ZBPs. In the case of the coupled chains we additio\-nal\-ly consider a finite phase difference in the order parameters of the SCs, yielding
$4\pi$-periodic Josephson MF couplings \cite{JosephsonEffect}, which can further control the location of the peaks. This collection of fingerprints appear feasible to be
looked for in existing platforms \cite{Yazdani_Science} and capable of unveiling the presence of MFs.

\section{Theoretical model and methods}

In the following section we present in detail our theoretical approach and model. In Sec.~\ref{Sec:Shiba} we review the topological properties and model of a single
ferromagnetic YSR chain. The form of the related MF wavefunctions and the inter-MF coupling in the case of two tunnel-coupled TSCs are both discussed in Sec.~\ref{Sec:MFs}.
The SPSTM tip Hamiltonian and its coupling to the MFs is contained in Secs.~\ref{Sec:TIP} and \ref{Sec:Tip-MF}. {\color{black}Finally, in Sec.~\ref{Sec:transport} we present
the steps for calculating the tunneling conductance by employing the Keldysh formalism, restricted to the Andreev processes induced by the MFs.}

\subsection{Topological ferromagnetic YSR chain}\label{Sec:Shiba}

In a previous work, Ref.~\onlinecite{Interplay}, three of the present authors extracted the microscopic topological model describing the YSR bands of a ferromagnetic chain
deposited on a superconducting substrate with Rashba SOC. The YSR bands arise from the overlap of YSR bound states which lie energetically deep below the ener\-gy scale 
set by the bulk superconducting gap. For extracting this low energy model, one starts from the Hamiltonian of the bulk SC with Rashba SOC
\bea
H_{\rm SC}=\frac{1}{2} \sum_{\bm k} \Psi_{\bm k}^\dag\left[\xi_{\bm{k}} \tau_z+\alpha \tau_z(\bm k \times \hat{\bm{z}})\cdot \tilde{\bm\sigma}-\Delta\tau_y\sigma_y\right]
\Psi_{\bm k}\,,\qquad\label{eq::Ham_supercond}
\eea

\noi where the Pauli matrices $\bm{\sigma}$ and $\bm{\tau}$ are defined in spin and particle-hole space respectively,
$\tilde{\bm{\sigma}}=(\tau_z\sigma_x,\sigma_y,\tau_z\sigma_z)$, while $\Psi_{\bm k}^\dag = (\psi_{\bm k\uparrow}^\dag,\,\psi_{\bm k\downarrow}^\dag,\,\psi^{\phantom\dag}_{-\bm
k\uparrow},\,\psi^{\phantom\dag}_{-\bm k\downarrow})$ is the correspon\-ding spinor. The operators $\psi_{\bm{k},\sigma}^{\dag}$ create electrons of momentum $\bm{k}$ and
spin-projection $\sigma$. Here $\xi_{\bm{k}}$ corresponds to the free electron energy dispersion, $\alpha$ denotes the SOC strength and $\Delta$ the bulk superconducting gap.
Straightforward manipulations allow us to obtain a Hamiltonian describing the electrons of the SC, localised at the adatom sites. The corresponding Schr\"odinger equation in
site space $i,j$ has the form (see Ref.~\onlinecite{Interplay})
\begin{align}
\label{eq::Sch}
\sum_j \widehat{{\cal H}}_{ij}\phi_j = {\varepsilon}\phi_i
\end{align}

\noi with $\phi_i^{\dag}=(u_{i,\uparrow}^*\,,u_{i,\downarrow}^*\,,v_{i,\uparrow}\,,v_{i,\downarrow})$ and the Hamiltonian
\bea
\label{eq::Hamiltonian_FM}
\widehat{{\cal H}}_{ij}=\frac{\Delta}{\pi\nu_F M^2}\bigg[\left(\pi\nu_FM^2\tau_y\sigma_y-M\tau_z\sigma_z\right)\delta_{ij}\qquad\qquad\phd\no\\
+M^2\left(G^s_{i-j}\tau_z-G^a_{i-j}\tau_z\sigma_y+F^s_{i-j}\tau_y\sigma_y-F^a_{i-j}\tau_y\right)\bigg]\,,\quad
\eea

\noi with the Fermi-level density of states (DOS) in the normal phase of the SC, $\nu_F$. $M=JS$ denotes the ferromagnetic energy scale, with $J$ the exchange energy
between the SC electrons and the adatoms with spin $S$. The solution of Eq.~\eqref{eq::Sch} determines the energies and wavefunctions of the YSR midgap states,
{\color{black}while the coefficients read}
\bea
\frac{G^s(r)}{\pi\nu_F}&=&\cos(k_F\alpha r/v_F)\sin\big(k_F|r|-\tfrac{\pi}{4}\big){e^{-\tfrac{|r|}{\xi_0}}}\sqrt{\tfrac{2}{\pi k_F|r|}}\,,\phd\quad\\
\frac{F^s(r)}{\pi\nu_F}&=&\cos(k_F\alpha r/v_F)\cos\big(k_F|r|-\tfrac{\pi}{4}\big){e^{-\tfrac{|r|}{\xi_0}}}\sqrt{\tfrac{2}{\pi k_F|r|}}\,,\phd\quad\\
\frac{G^a(r)}{i\pi\nu_F}&=&\sin(k_F\alpha r/v_F)\sin\big(k_F|r|-\tfrac{\pi}{4}\big){e^{-\tfrac{|r|}{\xi_0}}}\sqrt{\tfrac{2}{\pi k_F|r|}}\,,\phd\quad\\
\frac{F^a(r)}{i\pi\nu_F}&=&\sin(k_F\alpha r/v_F)\cos\big(k_F |r|-\tfrac{\pi}{4}\big){e^{-\tfrac{|r|}{\xi_0}}}\sqrt{\tfrac{2}{\pi k_F|r|}}\,,\phd\quad
\label{eq::GsGaFsFa}
\eea

\noi where $\xi_0$ is the coherence length of the SC, $k_F$ ($v_F$) the Fermi wave-vector (velocity) and $r=(i-j)a$, with the adatom spacing $a$. Note that $G^{a,s}(0) =
F^{a,s}(0) = 0$. The indices $s$ and $a$ denote functions which are symmetric or anti-symmetric under inversion $r\rightarrow-r$. In the rest of the manuscript we use $k_F$
in units of $\pi/a$, $\alpha$ in units of $v_F$, $M$ in units of $1/(\pi\nu_F)$ and $\xi_0$ in units of $a$.

By transferring to $k$-space we obtain the Bogoliubov-de Gennes (BdG) Hamiltonian
\bea\label{eq::HFMk}
\widehat{{\cal H}}_k=t_k\tau_z-v_k\tau_z\sigma_y+(\Delta+{\cal D}_k)\tau_y\sigma_y-d_k\tau_y-{\cal B}\tau_z\sigma_z\,,\quad\ph
\eea

\noi where we have introduced ${\cal B}=\Delta/(\pi\nu_F JS)$ and
\bea
t_k&=&\sum_{\delta=1}^{\infty}t_{\delta}\cos(\delta ka)\quad{\rm with}\quad t_{\delta}=\frac{2\Delta}{\pi\nu_F}\ph G^s_{\delta}\,,\\
v_k&=&\sum_{\delta=1}^{\infty}v_{\delta}\sin(\delta ka)\quad{\rm with}\quad v_{\delta}=\frac{2\Delta}{i\pi\nu_F}\ph G^a_{\delta}\,,\\
{\cal D}_k&=&\sum_{\delta=1}^{\infty}{\cal D}_{\delta}\cos(\delta ka)\;\;{\rm with}\quad{\cal D}_{\delta}=\frac{2\Delta}{\pi\nu_F}\ph F^s_{\delta}\,,\\
d_k&=&\sum_{\delta=1}^{\infty}d_{\delta}\sin(\delta ka)\quad{\rm with}\quad d_{\delta}=\frac{2\Delta}{i\pi\nu_F}\ph F^a_{\delta}\,.\qquad
\eea

The BdG Hamiltonian above resides in symmetry class BDI \cite{Interplay,Kotetes}, with time-reversal symmetry $\Theta={\cal K}$ ($\Theta^2=I$), chiral symmetry
$\Pi=\tau_x$ and charge-conjugation $\Xi=\tau_x{\cal K}$, where ${\cal K}$ denotes complex conjugation. The particular symmetry class supports a $\mathbb{Z}$ topological
invariant \cite{tenfold} allowing an integer number of MFs per chain edge. The detailed diagram of 0, 1 and 2 MF phases per edge has been extracted in
Ref.~\onlinecite{Interplay} as also in others works within different frameworks \cite{MacDonald,Hui}. To\-po\-lo\-gi\-cal YSR chains with multiple MFs edge modes can be found
also in two-dimensional systems \cite{Ojanen2d}.

External perturbations which violate chiral symmetry $\Pi$ (or equivalently $\Theta$) enforce the system to reside in symmetry class D, which allows up to a single MF per
edge. This implies that if the TSC resides in the topological phase with 2 MFs per edge, the application of an \textit{infinitesimally weak} $\Pi$-violating perturbation $m$,
will unavoidably hybridise the 2 MFs, splitting them into finite energy (proportional to $m$) bound states. However, the 1 MF per edge phase remains unaffected by such a
weak field. Symmetry analysis demonstrates that the simplest source of chiral symmetry breaking is a Zeeman field applied along the $y$ axis, i.e. $B_y$. Instead $B_x$ and
$B_z$ fields preserve the latter. The $B_y$ field can be either applied globally or only near the edge, since it will always hybridise the 2 MFs sitting at the same edge.
Notably in a SPSTM experiment the tip itself is magnetised, and depending on its magnetic orientation it can controllably violate or preserve $\Pi$. The latter property has
significant ramifications when measuring the tunneling conductance with the SPSTM technique, and can influence the possible observation of the ZBP or its quantisation with
a single or double unit of conductance ($2e^2/h$).

\subsection{Majorana wavefunctions and their coupling}\label{Sec:MFs}

By solving Eq.~\eqref{eq::Sch} for a {\color{black}finite-size} chain, one can retrieve the MF wavefunctions which have the special form
\bea
\Phi_{i,n}^{\dag}=\left(u_{i,\uparrow,n}^*\,,u_{i,\downarrow,n}^*\,,u_{i,\uparrow,n}\,,u_{i,\downarrow,n}\right)\,,
\eea

\noi with the wavefunction components satisfying $\sum_{i,\sigma}|u_{i,\sigma,n}|^2=1/2$. The latter normalisation implies the anticommutation relation for the corresponding
MF operators $\{\gamma_n,\gamma_m\}=\delta_{n,m}$. For a low energy description, one can focus only on the MF sector and neglect the rest of the BdG quasiparticles which lie
above the bulk superconducting gap. Thus in the low energy limit, we can approximate the YSR state operators with the MF operators, i.e.,
$\psi_{i,\sigma}=\sum_nu_{i,\sigma,n}\gamma_n$. In the presence of suitable symmetries and ideal conditions (e.g. infinite chains) the MFs operators remain unpaired
\cite{KitaevUnpaired}, while in any realistic situation one has to introduce also couplings for the MFs, {\color{black}yielding the following general Hamiltonian}
\bea
{\cal H}_{\rm MF}=\frac{\imath}{2}\sum_{n,m}{\cal M}_{nm}\gamma_n\gamma_m\,.
\eea

Solely residing on the above MF description is a good approximation only as long as the coupling elements ${\cal M}_{nm}$ are much smaller compared to the YSR bandstructure
gap. As long as this is case, the matrix elements can arise due to the following reasons: i) a weak chiral symmetry breaking term, e.g. $m$, coupling for instance 2 MFs at the
same edge previously protected by $\Pi$, ii) finite size of the chains which can allow the overlap of MFs prima\-ri\-ly located at the far edges, giving rise to quasiparticles
with finite energy splitting $\delta\epsilon$ and iii) coupling $M$ of neighbouring MFs located at the edge of two different tunnel-coupled TSC chains. For the purposes of our
discussion MF coupling matrix elements of the first type will be discussed at a phenomeno\-lo\-gi\-cal level, while for the others we will explicitly calculate the
couplings by employing the BdG formalism and considering a particular model for inter-chain tunneling, respectively.

For inferring the coupling of edge MFs due to the inter-chain tunneling, we focus on the electronic degrees of the two substrate SCs and consider that they are separated
by a thin insulating film yielding a Josephson junction with a corresponding superconducting phase difference $\delta\varphi$. The latter can be imposed by inducing a
supercurrent flow through the junction or by gluing together the very left and right edges of a single chain in order to form a ring through which we can thread flux. To this
end, we assume that the electrons of the two substrate SCs are coupled via the Hamiltonian
\bea
{\cal H}_{\rm T}=\sum_{i,j}\left[\psi_{i,\sigma}^{\dag}T_{i,j}e^{\imath(\varphi_i-\varphi_j)/2}\psi_{j,\sigma}+{\rm H.c.}\right]\,.
\eea

\noi In particular, here we consider the profile
\bea
T_{i,j}=t\frac{1-{\rm sgn}(i\cdot j)}{2}\exp\left[-\frac{|ai-bj|-(a+b)}{l}\right]\,.\quad
\eea

\noi In the above we have already assumed that one TSC chain resides on the positive sites $i,j>0$ and the other on the negative ones. In this manner, the projector
$[1-{\rm sgn}(i\cdot j)]/2$ allows only interchain tunneling, while we have also assumed that the tunneling strength decays exponentially with the distance $|ia-jb|$ over a
characte\-ri\-stic decay length $l$, with $a,b$ denoting the adatom spa\-cings of each chain. Moreover, in the above Hamiltonian we have incorporated the spatially varying
supercon\-duc\-ting phase profile $\varphi_i$, which here is assumed to have the form $\varphi_i={\rm sgn}(i)\delta\varphi/2$. The corresponding MF matrix elements that one
obtains take the following form
\bea
M_{nm}=4 \Im\sum_{i,j,\sigma} u^*_{i,\sigma,n}T_{i,j}e^{\imath(\varphi_i-\varphi_j)/2}u_{j,\sigma,m}\,.
\eea

\subsection{SPSTM tip Hamiltonian}\label{Sec:TIP}

For the purposes of this work we model the SPSTM tip at site $i$, as a lead of spinful electrons under the influence of a spin splitting field $\bm{P}$. Moreover, here we
assume that the tip feels a voltage $V_i$, which is responsible for driving the coupled system out of equilibrium and leads to the tunneling current. Previous works on the
tunneling conductance for such systems have considered a non-magnetic tip in the normal \cite{Peng,ZBPDSarma} or the superconducting phase \cite{SCtip}. The tip Hamiltonian
has the form
\bea
{\cal H}_{{\rm TIP},i}=
\sum_{\bm{k},\alpha,\beta}[(\epsilon_{\bm{k}}-eV_i)\delta_{\alpha\beta}-\bm{P}\cdot\bm{\sigma}_{\alpha\beta}]c_{\bm{k},\alpha,i}^{\dag}c^{\phantom \dag}_{\bm{k},\beta,i}
\,.\quad
\eea

\noi Here by keeping the index $i$ for the tip electrons and the voltage we allow our formalism to address the case of a multi-tip SPSTM, while in the most general case we
should include an index $i$ also to the spin-polarisation $\bm{P}$. The latter can be parametrised in the following way
\bea
\bm{P}=P(\cos\vartheta\sin\eta,\sin\vartheta\sin\eta,\cos\eta)\,.
\eea

\noi The presence of $\bm{P}$ modifies the DOS for the spin up and down electrons, $\rho_{\sigma}(E)$, at energy $E$. In particular the normalised DOS read
\bea
\nu_{\sigma}=\rho_{\sigma}(E)/\rho(E)\quad{\rm with}\quad \rho(E)=\sum_{\sigma}\rho_{\sigma}(E)\,,\quad
\eea

\noi {\color{black}where $\rho_{\uparrow}-\rho_{\downarrow}\propto P$ and $\sum_{\sigma}\nu_{\sigma}=1$. We now introduce the spin polarisation degree
$P_s=\nu_{\uparrow}-\nu_{\downarrow}$, ranging from $-1$ to $1$. The latter values occur for fully spin-polarised tips. $P_s$ can strongly vary depending on the tip material,
while it is in principle possible to achieve complete polarisation ($P_s=\pm1$) by fabricating the tip using a half-metal, for which one of the spin bands does not cross the
Fermi level (for more details see Ref.~\onlinecite{Wiesendanger}).}

{\color{black}For the rest of our analysis, it is eligible to perform a rotation and diagonalise the tip Hamiltonian in spin space using the property}
\bea
\bm{P}\cdot\bm{\sigma}=\widehat{R}^{\dag}P\sigma_z\widehat{R}\,,\quad\widehat{R}=\exp(\imath\eta\sigma_y/2)\exp(\imath\vartheta\sigma_z/2)\,.\qquad
\eea

\noi Thus the tip Hamiltonian becomes
\bea
{\cal H}_{{\rm TIP},i}=\sum_{\bm{k},\sigma=\pm}(\epsilon_{\bm{k},\sigma}-eV_i)\tilde{c}_{\bm{k},\sigma,i}^{\dag}\tilde{c}^{\phantom\dag}_{\bm{k},\sigma,i}\,.
\eea

\noi with $\sigma=\pm$ labelling the two eigenstates of $\sigma_z$ with eigenvalues $\pm 1$ and $\epsilon_{\bm{k},\sigma}=\epsilon_{\bm{k}}-\sigma P$.

\subsection{Coupling between the SPSTM tip and the MFs}\label{Sec:Tip-MF}

The SPSTM tip originally couples to the electronic density of the superconducting substrate at site $i$, via the tunneling Hamiltonian
\bea
{\cal H}_{\text{TIP-MF},i}=\sum_{\bm{k},\sigma}\left({\rm T}_{\bm{k},i}c_{\bm{k},\sigma,i}^{\dag}\psi_{i,\sigma}^{\phantom\dag}+{\rm H.c.}\right)\,.
\eea

\noi After diagonalising the tip Hamiltonian in spin space we obtain
\bea
{\cal H}_{\text{TIP-MF},i}=\sum_{\bm{k},\sigma,n}
\left({\cal V}^{\phantom \dag}_{\bm{k},\sigma,i,n}\tilde{c}_{\bm{k},\sigma,i}^{\dag}-{\cal V}_{\bm{k},\sigma,i,n}^*\tilde{c}^{\phantom \dag}_{\bm{k},\sigma,i}\right)
\gamma_n\,,\qquad
\eea

\noi with the matrix elements
\bea
{\cal V}_{\bm{k},\sigma,i,n}=\sum_{\sigma'}{\rm T}_{\bm{k},i}R_{\sigma,\sigma'}u_{i,\sigma',n}\,.
\eea

\subsection{Tunneling conductance}\label{Sec:transport}

After having set the stage for calculating the tunneling conductance, we can proceed with deriving the Heisenberg operator for the current through the tip located over
site $i$
\bea
\hat{I}_i(t)&=&-e\dot{N}_{\rm
TIP}(t)=-e\sum_{\bm{k},\sigma}\frac{d}{dt}\left(\tilde{c}_{\bm{k},\sigma,i}^{\dag}(t)\tilde{c}^{\phantom\dag}_{\bm{k},\sigma,i}(t)\right)\no\\
&=&-\frac{e\imath}{\hbar}\sum_{\bm{k},\sigma}\left[{\cal H}_{\text{TIP-MF},i},\,\tilde{c}_{\bm{k},\sigma,i}^{\dag}\tilde{c}^{\phantom\dag}_{\bm{k},\sigma,i}\right](t)\,.
\eea

\noi The current operator in the Schr\"{o}dinger picture reads
\bea
\hat{I}_i=\frac{e\imath}{\hbar}\sum_{\bm{k},\sigma,n}
\left({\cal V}^{\phantom\dag}_{\bm{k},\sigma,i,n}\tilde{c}_{\bm{k},\sigma,i}^{\dag}+{\cal V}_{\bm{k},\sigma,i,n}^*\tilde{c}^{\phantom\dag}_{\bm{k},\sigma,i}\right)\gamma_n\,.
\eea

\noi The expectation value of the current operator $I_i(t)\equiv\langle\hat{I}_i(t) \rangle$ is calculated from the expression
\bea
I_i(t)&=&\frac{2e}{\hbar}\sum_{\bm{k},\sigma,n}\Im\left[{\cal V}_{\bm{k},\sigma,i,n}^*\left<\gamma_n(t)\tilde{c}_{\bm{k},\sigma,i}(t)\right>\right]\no\\
&=&-\frac{2e}{\hbar}\sum_{\bm{k},\sigma,n}\Re\left[{\cal V}_{\bm{k},\sigma,i,n}^*G^<_{\bm{k},\sigma,i,n}(t,t)\right]\,,
\eea

\noi which involves the lesser mixed Green's function
\bea
G^<_{\bm{k},\sigma,i,n}(t,t')\equiv \imath\left<\gamma_n(t')\tilde{c}_{\bm{k},\sigma,i}(t)\right>\,.
\eea

\noi For calculating the current one can employ the Keldysh formalism and introduce the respective Keldysh-contour-ordered Green's functions. After following this route we
find that for retrieving the final expression for the current we will need the retarded MF Green's functions
\bea
{\cal G}_{nm}^R(t,t')\equiv-\imath\Theta(t-t')\left<\gamma_n(t)\gamma_m(t')\right>\,.
\eea

\noi Since here we are interested in the non-equilibrium steady state, one can follow the method of Ref.~\onlinecite{FlensbergTunneling} and show that the expectation value
of the current operator is given by the formula
\bea
I_i=\frac{e}{h}\int\limits_{-\infty}^{+\infty}d\omega\, {\cal T}_i(\omega)\left[n_F(\omega-eV_i)-n_F(-\omega+eV_i)\right]\,,\quad\phd
\eea

\noi with the Fermi-Dirac distribution $n_F(\omega)$ at energy $\omega$ and the transmission coefficient
\bea
{\cal T}_i(\omega)\equiv\operatorname{Tr}\left[\widehat{{\cal G}}^R(\omega)\widehat{\Gamma}^{i*}(-\omega)\widehat{{\cal G}}^A(\omega)\widehat{\Gamma}^i(\omega)\right]\,,
\eea

\noi according to terminology of the Landauer-B\"uttiker formalism. Note however that in contrast to the usual Landauer formula for bal\-li\-stic trasport in normal
conductors, the present formula involves one electron and one hole Fermi-Dirac distribution. This is a direct consequence of the involvement of MFs which are
equal superpositions of electrons and holes. The matrix elements for the linewidth hermitian matrices $\widehat{\Gamma}^i(\omega)$ are given by the
expression
\bea
\Gamma_{nm}^i(\omega)=2\pi\sum_{\bm{k},\sigma}{\cal V}_{\bm{k},\sigma,i,n}^*{\cal V}_{\bm{k},\sigma,i,m}\delta(\omega-\epsilon_{\bm{k},\sigma})\,.
\eea

\noi As in Ref.~\onlinecite{FlensbergTunneling} we adopt the wideband approximation according to which the linewidth matrix elements can be considered to be energy
independent. In particular, we assume that ${\rm T}_{\bm{k},i}={\rm T}$ (the $\bm{k}$ independence reflects the wideband approximation) and set the DOS of the tip to be
approximately $\rho_{\sigma}(E_F)$. The latter consideration yields
\bea
\Gamma_{nm}^i(\omega)\equiv\Gamma_{nm}^i=\Gamma \bm{u}_{i,n}^{\dag}\frac{\mathds{1}+P_s\hat{\bm{P}}\cdot\bm{\sigma}}{2}\ph\bm{u}_{i,m}\,,
\eea

\noi with $\Gamma=2\pi\rho(E_F)|{\rm T}|^2$, $\bm{u}_{i,n}^{\dag}=(u_{i,\uparrow,n}^*,\, u_{i,\downarrow,n}^*)$ and the unit vector $\hat{\bm{P}}=\bm{P}/P$. Note that for
convenience we set $\Gamma=1$, a convention which we will follow throughout the remainder of the manuscript. For retrieving the transmission coefficent one has to calculate
the retarded and advanced MF matrix Green's functions given by
\bea
\widehat{{\cal G}}^R(\omega)=\left(\omega\hat{I}-\imath \widehat{{\cal M}}+\imath\Re\widehat{\Gamma}^i\right)^{-1}\,,
\eea

\noi where $\widehat{{\cal G}}^A(\omega)=[\widehat{{\cal G}}^R(\omega)]^{\dag}$. In the above we have con\-si\-de\-red that the self-energies of the MF matrix Green's
functions contain only the linewidth fun\-ctions, in the spirit of the wideband approximation as in Ref.~\onlinecite{FlensbergTunneling}.

For the rest of our discussion, we will focus on the \textit{zero temperature} tunneling conductance given by the fol\-lowing expression
\bea
\frac{dI_i}{dV_i}=\frac{2e^2}{h}{\cal T}_i(eV)\,.
\eea

\section{Results and discussion}

In the following paragraphs we present our results concerning the tunneling conductance for a single or two coupled TSC magnetic chains. In particular, the case of a single
chain is examined in Secs.~\ref{Sec:dIdV2d01} and \ref{Sec:dIdV2d02} where we analyse TSCs with 1 and 2 MFs per edge, respectively. Later on, we consider the situation of two
coupled TSC chains with the superconducting phases of the substrate SCs differing by $\delta\varphi$, thus allowing a 4$\pi$-periodic Josephson coupling between the
neighbouring edge MFs. Specifically, in Sec.~\ref{Sec:dIdV2d11} we consider two chains, each of which harbors a single MF per edge. In
Sec.~\ref{Sec:dIdV2d21}(\ref{Sec:dIdV2d12}) we investigate Josephson junctions consisting of a TSC chain with a single MF per edge and a TSC chain with 2 MFs per edge, where
we assume that the SPSTM tip couples to the chain hosting 1(2) MFs per edge. In Sec.~\ref{Sec:dIdV2d22} we considered a Josephson junction of two TSC chains each of which
hosts 2 MFs per edge. In the whole discussion particular emphasis will be given to chiral symmetry violation and restoration, and its implications for observing the ZBPs which
are considered smoking gun signatures of MFs.

\subsection{One TSC chain with 1 MF per edge}\label{Sec:dIdV2d01}

\begin{figure}[t]
\centering
\includegraphics[scale=0.25]{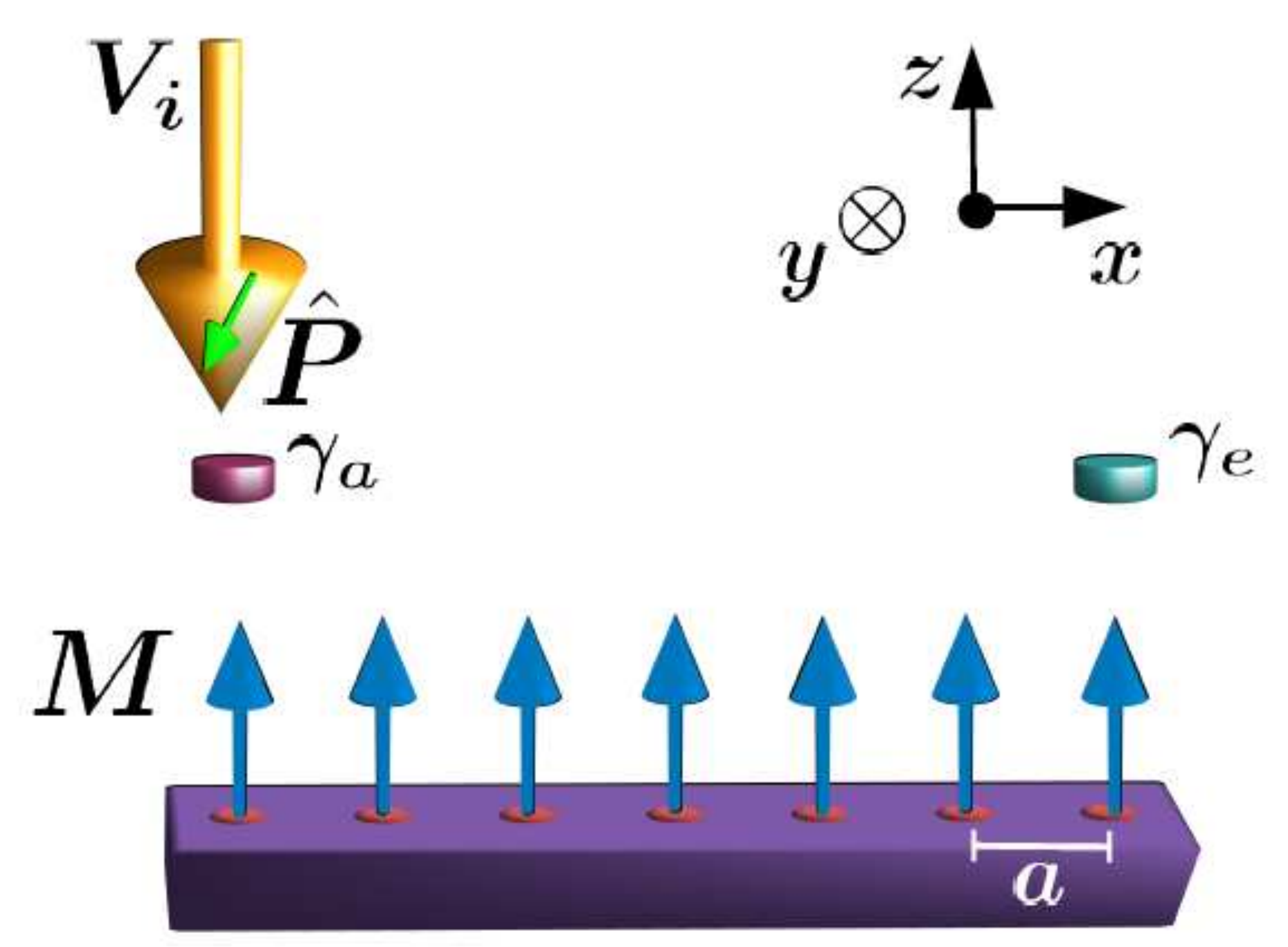}
\vspace{0.1in}
\caption{TSC based on a ferromagnetically ordered atomic chain probed by SPSTM. The tip is considered to couple locally, i.e. only to the electronic density directly 
below it. For long chains only the $\gamma_a$ MF is seen by the probe, while for short ones both MFs contribute.}
\label{fig:dIdV2d01}
\end{figure}

In the present paragraph we discuss the case of an isolated chain with a single MF per edge as shown in Fig.~\ref{fig:dIdV2d01}. Similarly to the result of
Ref.~\onlinecite{FlensbergTunneling} the tunneling conductance in this case takes the simple Lorentzian form
\bea
\frac{dI_i}{dV}=\frac{2e^2}{h}\frac{(\Gamma_{aa}^i)^2}{(eV)^2+(\Gamma_{aa}^i)^2}\,,
\eea

\noi from which one recovers the ZBP with one quantum of conductance. In the above and from now on, we set $V_i\equiv V$ for convenience. The broadening is given by the
expression
\bea
\Gamma_{aa}^i=\Gamma \bm{u}_{i,a}^{\dag}\frac{\mathds{1} + P_s\hat{\bm{P}}\cdot\bm{\sigma}}{2}\ph\bm{u}_{i,a}\,.
\eea

\begin{figure}[t]
\centering
\includegraphics[width=0.49\columnwidth]{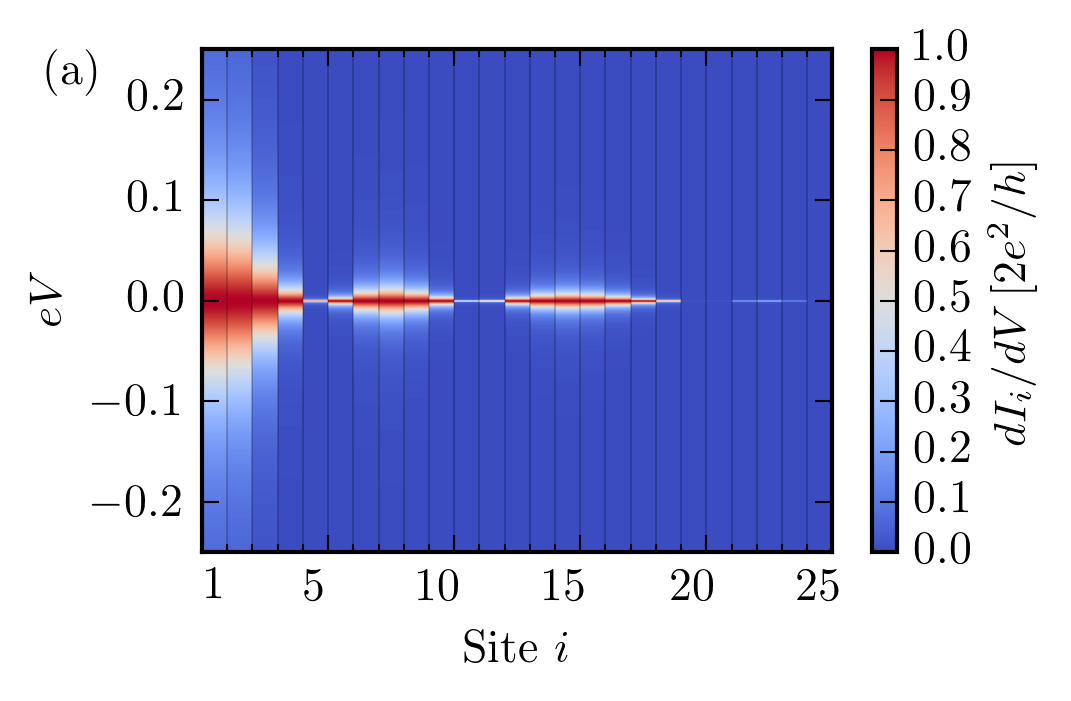}
\includegraphics[width=0.49\columnwidth]{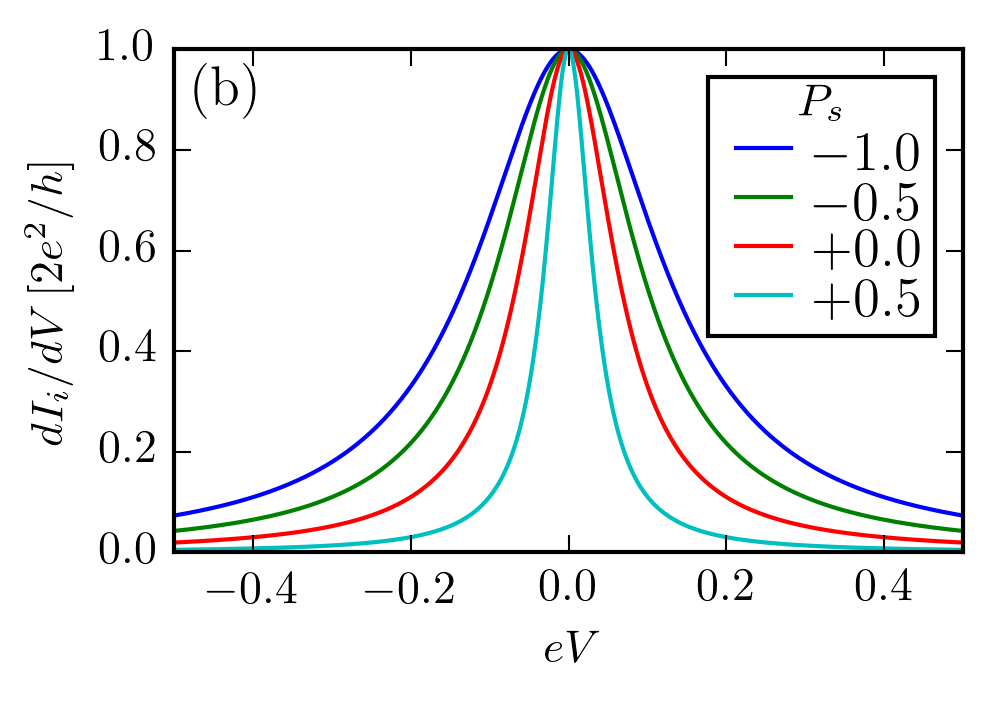}
\vspace{0.1in}
\includegraphics[width=1\columnwidth]{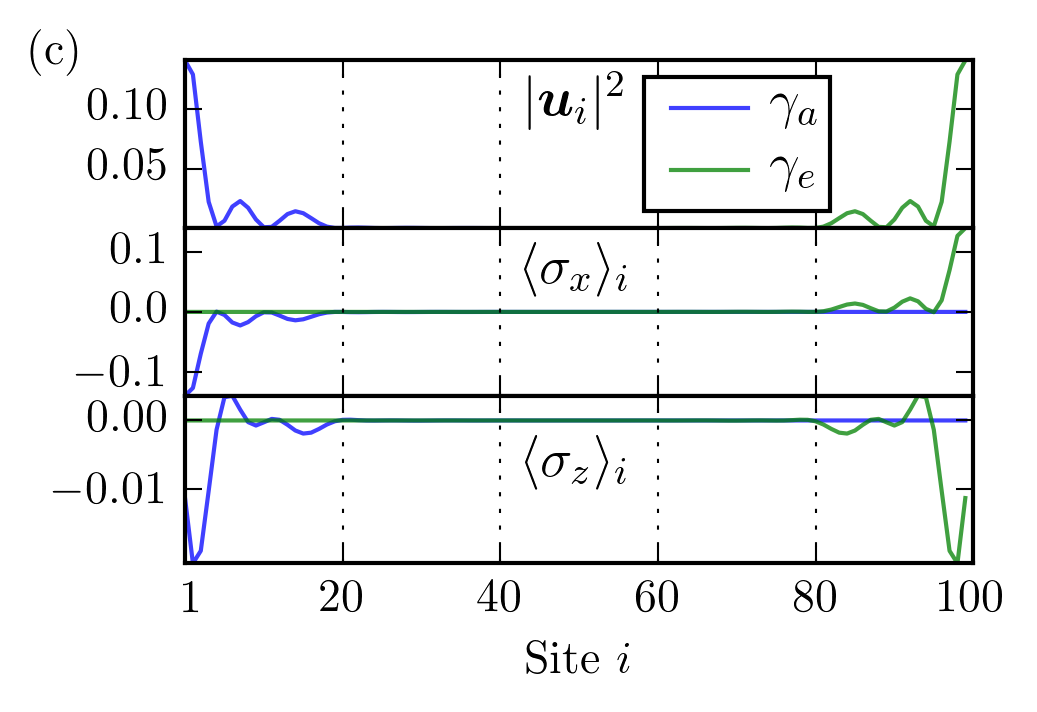}
\vspace{0.1in}
\includegraphics[width=0.49\columnwidth]{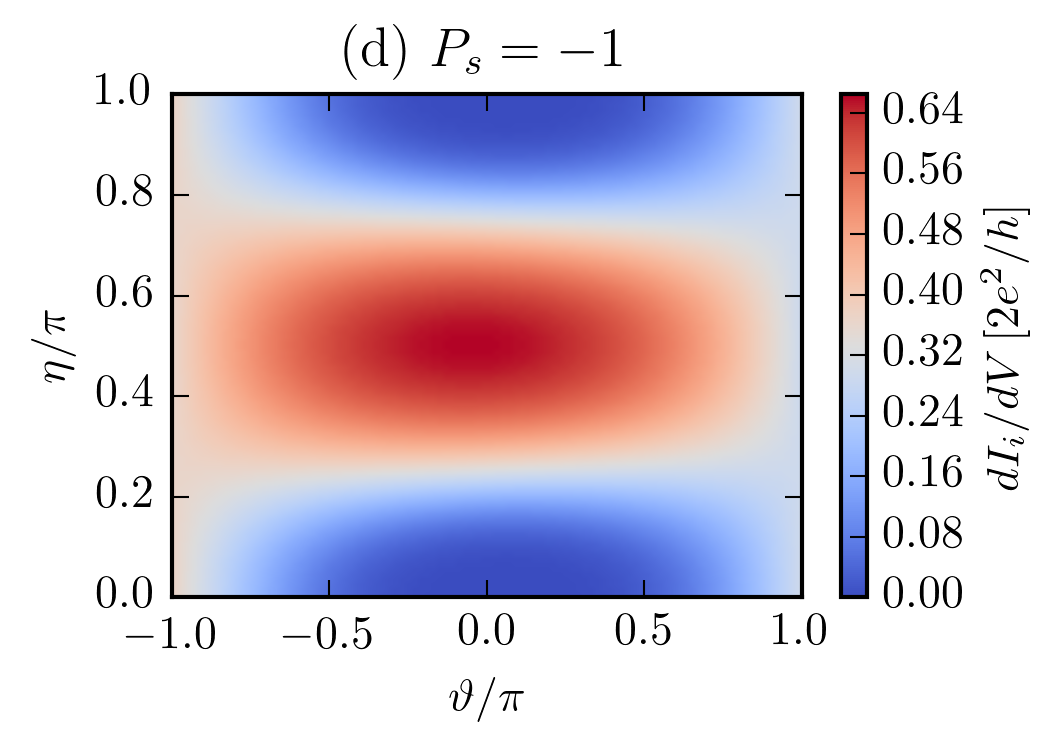}
\includegraphics[width=0.49\columnwidth]{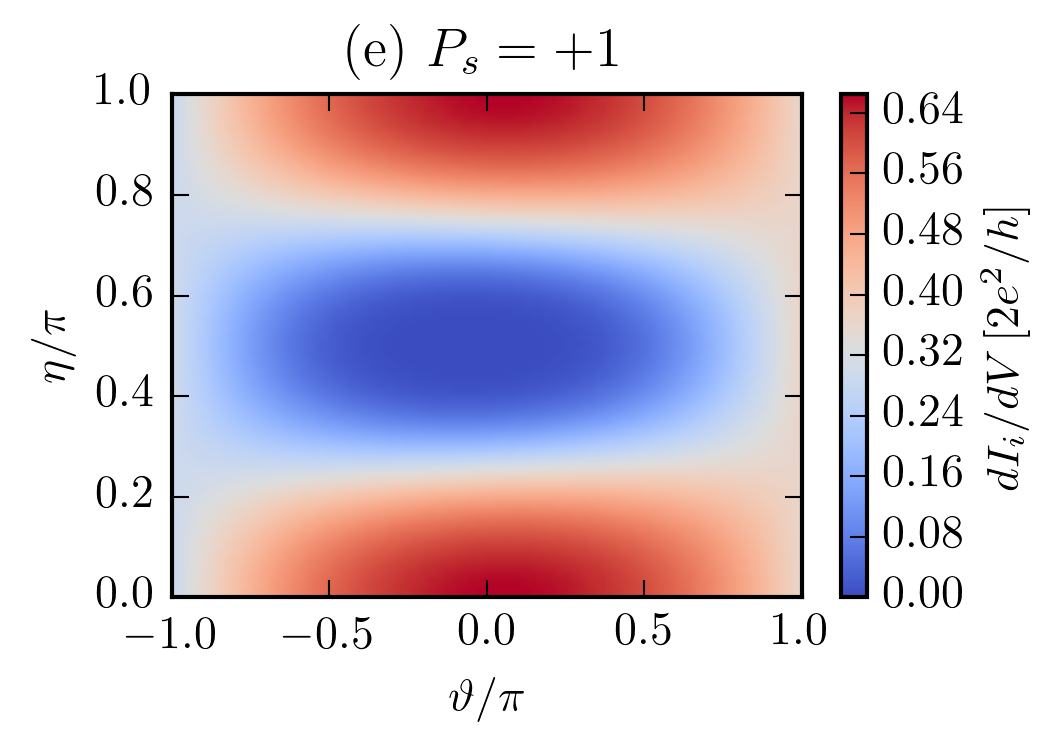}
\caption{SPSTM tip coupled to a single MF for the case of a sufficiently long chain of $N=100$ sites. (a) Spatial profile of the tunneling conductance for a spin unpolarised
tip ($P_s=0$). (b) {\color{black}Tunneling conductance at site $i=1$} for a tip polarised along the $x$ axis for varying $P_s$. The modification of the polarisation degree
alters the profile broadening. (c) Local spectral weight ($|\bm{u}_{i}|^2$) and electronic spin polarisation ($\langle\bm{\sigma}\rangle$) of the two MFs. Note that the spin
polarisation along the $y$ axis is exactly \textit{zero} as a result of $\Theta={\cal K}$ symmetry. (d, e) {\color{black}Tunneling conductance at site $i=1$} as a function of
the angles $(\vartheta,\eta)$ for the case of a fully polarised tip $P_s=\pm1$. The present maps {\color{black}reveal} the strong electronic spin anisotropy induced by the MF
$\gamma_a$ and constitutes one of its characteristic signatures. Parameters: $\xi_0=80$, $k_F=6.0$, $\alpha=0.01$ and $M=0.85$.}
\label{fig:dIdVPanel1MF}
\end{figure}

Crucial feature of our analysis is the inclusion of the magnetic characteristics of the SPSTM tip which opens new perspectives for detecting MFs. In fact, one observes that if
$\hat{\bm{P}}\cdot\bm{\sigma}\ph\bm{u}_{i,a}=-\bm{u}_{i,a}$ the linewidth term becomes $\Gamma_{aa}^i=\nu_{\downarrow}\Gamma\bm{u}_{i,a}^{\dag}\bm{u}_{i,a}$. If the tip
becomes fully spin polarised, so that $\nu_{\downarrow}=0$, the tunneling conductance will also go to zero and the ZBP will disappear from the tunneling spectra.
Essentially, when these conditions are met, the spin polarisation of the tip-electrons is anti\-pa\-ral\-lel to the electronic edge polarisation
$\left<\bm{\sigma}\right>_i=\bm{u}_{i,a}^{\dag}\bm{\sigma}\ph\bm{u}_{i,a}$ induced by the MF. Therefore the tip-MF coupling becomes zero since tunneling between the tip and
the substrate electrons is spin conserving and cannot take place for antiparallel polarisations.

As shown in Fig.~\ref{fig:dIdVPanel1MF} the MFs of both sides of a single chain induce an electronic spin-polarisation which is confined in the $xz$ plane, as a result of the
complex conjugation symmetry $\Theta={\cal K}$ which forces the spin-part of the wavefunction to be real. {\color{black}Note that a similar spin polarisation profile was
previously retrieved for nanowire-based TSCs in Ref.~\onlinecite{MajoranaSpinPolarisation}, as a result of the common features of the two models. The particular
distinctive feature, allows us to employ a SPSTM tip for unvei\-ling the MFs. In particular, one obtains a characteristic anisotropic dependence of the tunneling conductance
on the angles $(\vartheta,\eta)$ which determine the orientation of the tip-magnetisation, as it has been also pointed out in Ref.~\onlinecite{SESARS} within a different
context.}

In order to make a connection to the realistic situation, potentially relevant to the experiment of Ref.~\onlinecite{Yazdani_Science}, we further take into account the
influence of the remaining MF away from the tip. If the overlap of the MF wavefunctions is non-negligible due to the short length of the chain, a finite coupling of the form
$\imath\delta\epsilon\gamma_a\gamma_e$ will appear leading to finite energy excitations. More importantly, for short chains the tip generally couples to both MFs. In this case
we have
\bea
\widehat{{\cal M}}=\left(\begin{array}{cc}0&\delta\epsilon\\-\delta\epsilon&0\\\end{array}\right)\quad{\rm and}\quad
\widehat{\Gamma}^i=\left(\begin{array}{cc}\Gamma_{aa}^i&\Gamma_{ae}^i\\(\Gamma_{ae}^i)^*&\Gamma_{ee}^i\\\end{array}\right)\,,\quad\label{eq:MG01}
\eea

\begin{figure}[t]
\centering
\includegraphics[width=0.97\columnwidth]{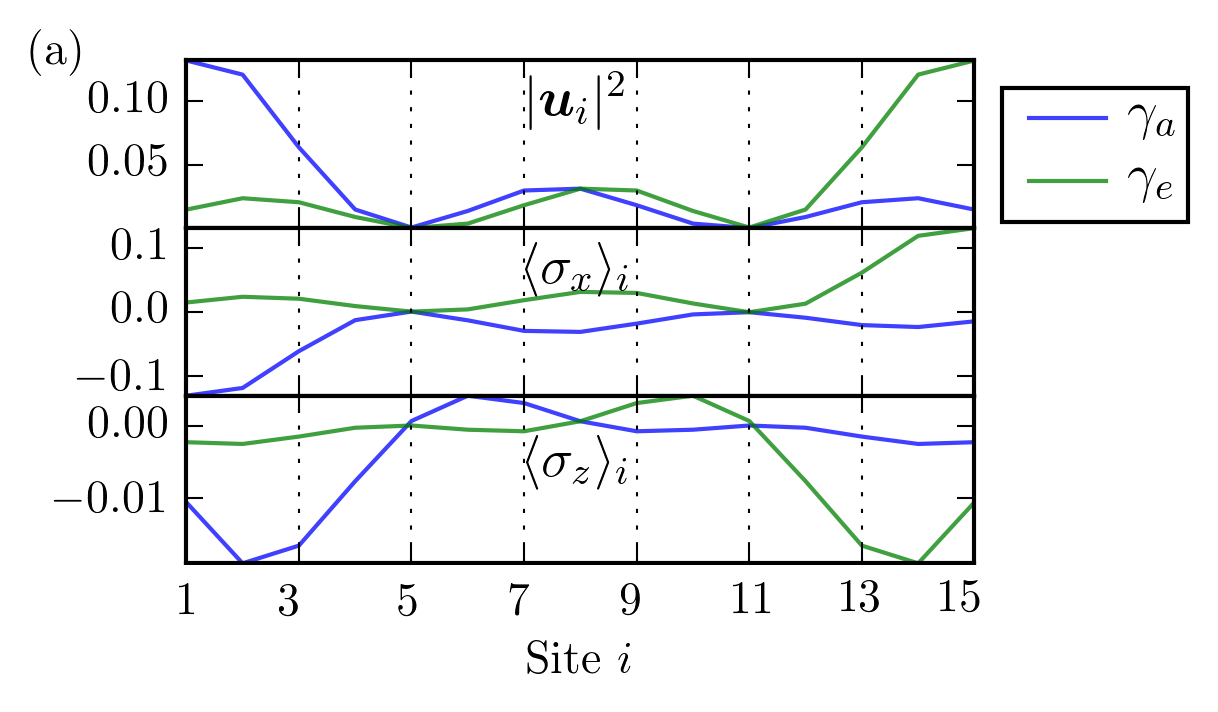}\qquad
\vspace{0.1in}
\includegraphics[width=1\columnwidth]{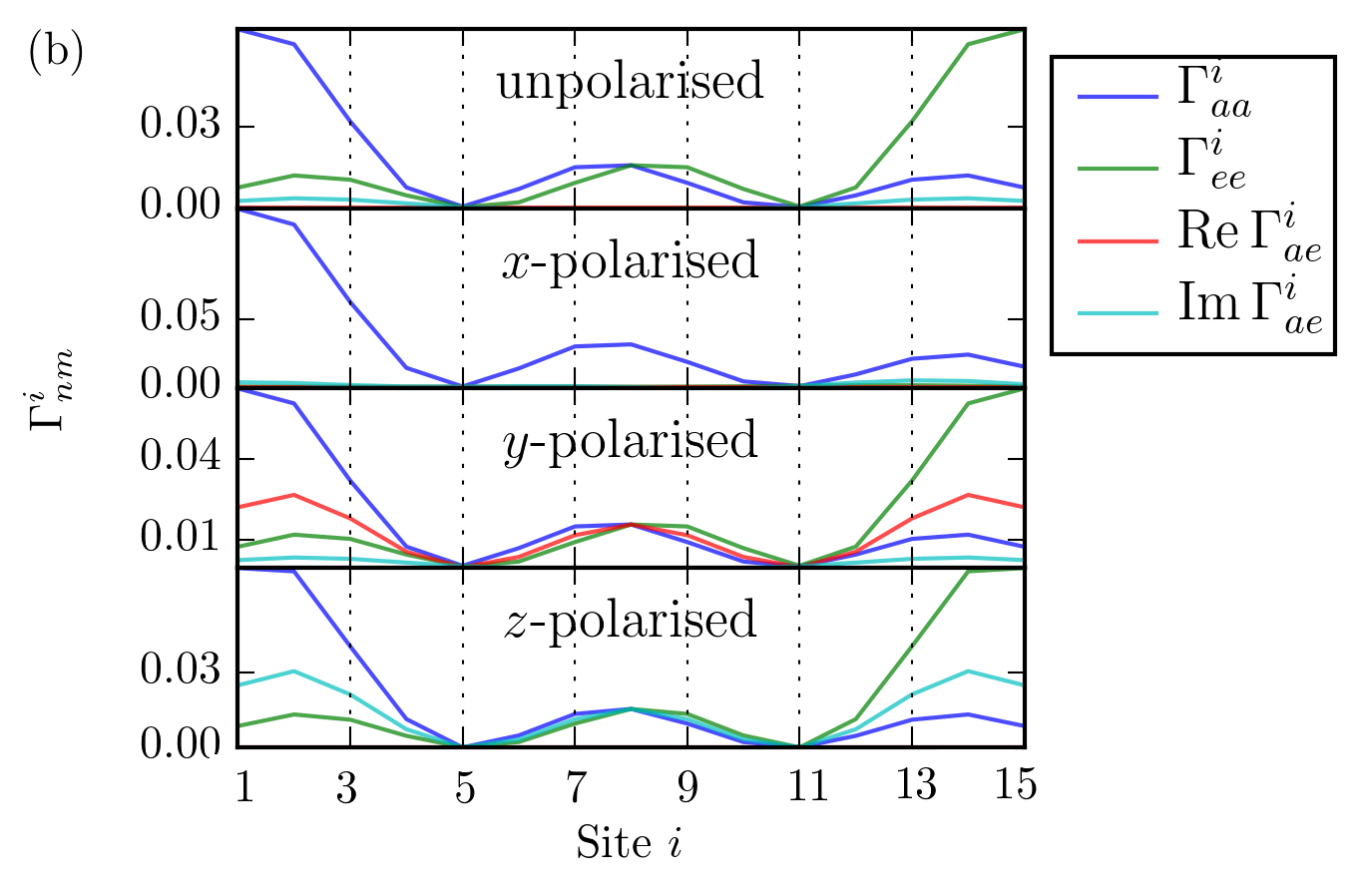}
\caption{(a) Spectral weight of MF wavefunctions and their induced electronic spin-polarisation in the case of a short chain consisting of $N=15$ sites. The two MFs
hybridise and form finite energy quasiparticle excitations. (b) Matrix elements of the linewidth matrix $\widehat{\Gamma}^i$. For each case {\color{black}corresponding to a
fully spin-polarised tip}, we present only the non-zero elements. Pa\-ra\-me\-ters: $\xi_0=80$, $k_F=6.0$, $\alpha=0.01$ and $M=0.85$.}
\label{fig:dIdVPanel1SingleMFshort}
\end{figure}

\noi yielding the modified tunneling conductance formula
\bea
&&\frac{dI_i}{dV}=\frac{2e^2}{h}\left\{2\left[\det(\Re\widehat{\Gamma}^i)+\delta\epsilon^2\right]\det\widehat{\Gamma}^i\right.\\
&&\left.+(eV)^2\left[\left(\Gamma_{aa}^i\right)^2+\left(\Gamma_{ee}^i\right)^2+2\left(\Re^2\Gamma_{ae}^i-\Im^2\Gamma_{ae}^i\right)\right]\right\}\cdot\no\\
&&\left\{\left[(eV)^2-\delta\epsilon^2-\det(\Re\widehat{\Gamma}^i)\right]^2
+(eV)^2\left(\Gamma_{aa}^i+\Gamma_{ee}^i\right)^2\right\}^{-1},\no\label{eq:dIdV01}
\eea

\noi For $V=0$ one obtains the simple result
\bea
\left.\frac{dI_i}{dV}\right|_{V=0}=2\cdot\frac{2e^2}{h}
\frac{\Gamma_{aa}^i\Gamma_{ee}^i-|\Gamma_{ae}^i|^2}{\Gamma_{aa}^i\Gamma_{ee}^i-\Re^2\Gamma_{ae}^i+\delta\epsilon^2}\,,
\eea

\noi which provides the ZBP height in this general case. Stri\-kin\-gly, when both MFs are accessed by the tip, the ZBP \textit{persists}, though with a conductance with
reduced spectral weight from the ideal value. This ZBP appears due the coupling of the tip to the $\gamma_e$ MF, and directly disappears if we set
$\Gamma_{ee}^i=\Gamma_{ae}^i=0$. Apart from the residual spectral weight for $V=0$, the conductance shows finite bias peaks
\bea
eV=\pm\sqrt{\delta\epsilon^2+\det(\Re\widehat{\Gamma}^i)}\,,
\eea

\begin{figure}[t]
\centering
\includegraphics[width=0.46\columnwidth]{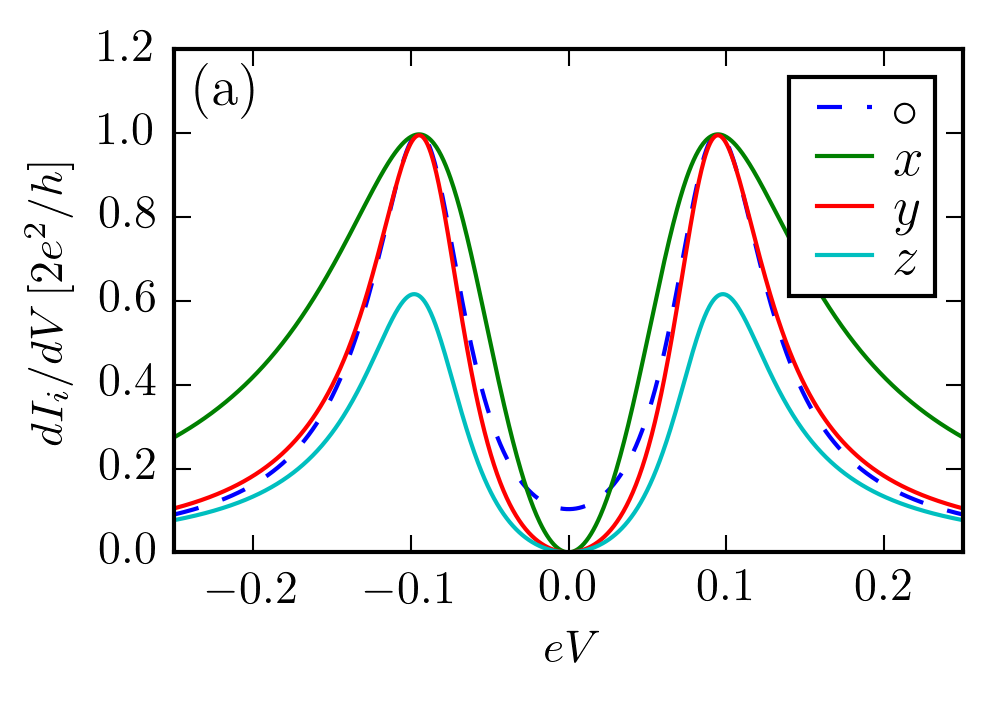}
\includegraphics[width=0.49\columnwidth]{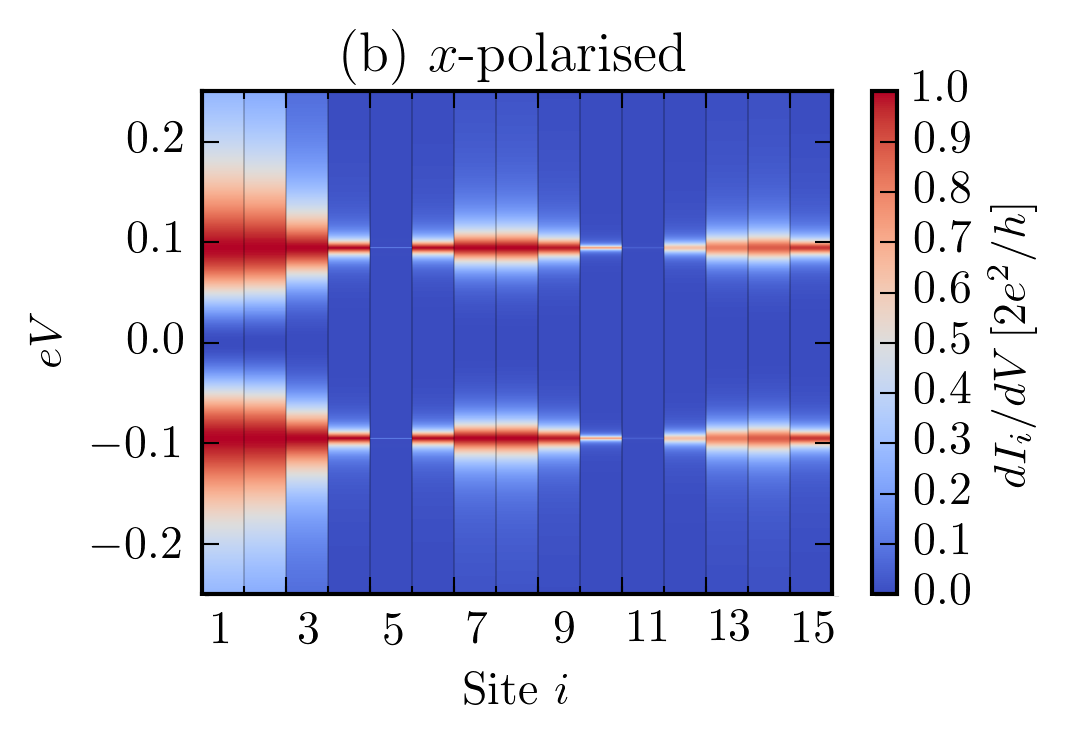}
\vspace{0.1in}
\includegraphics[width=0.49\columnwidth]{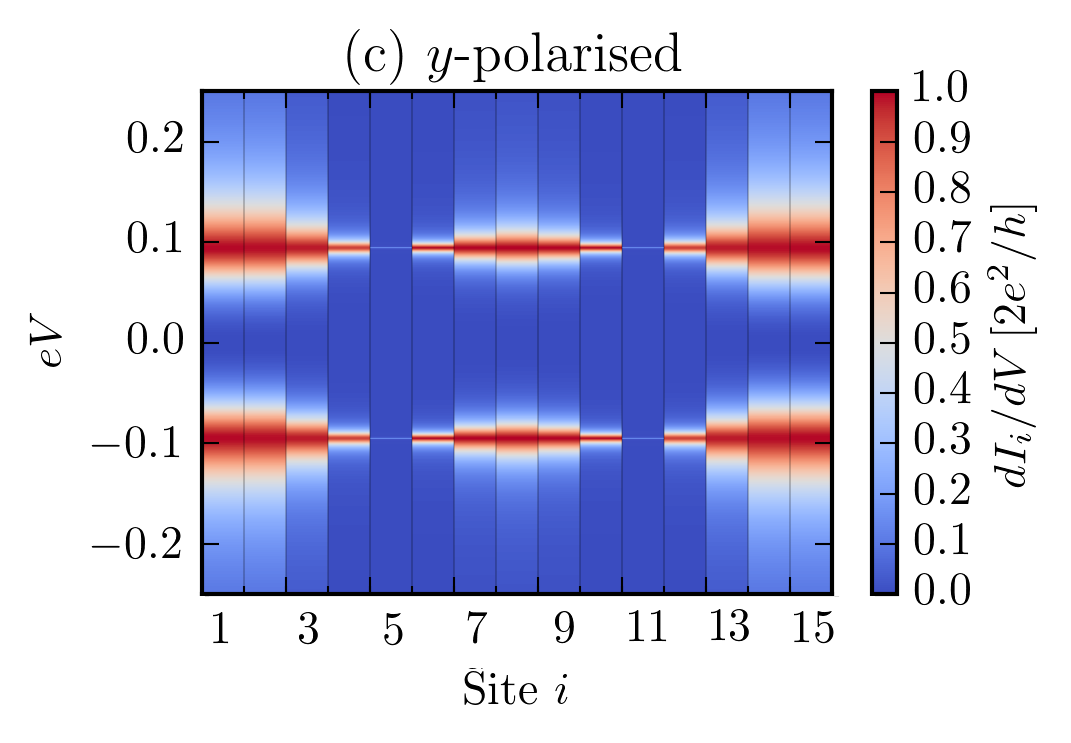}
\includegraphics[width=0.48\columnwidth]{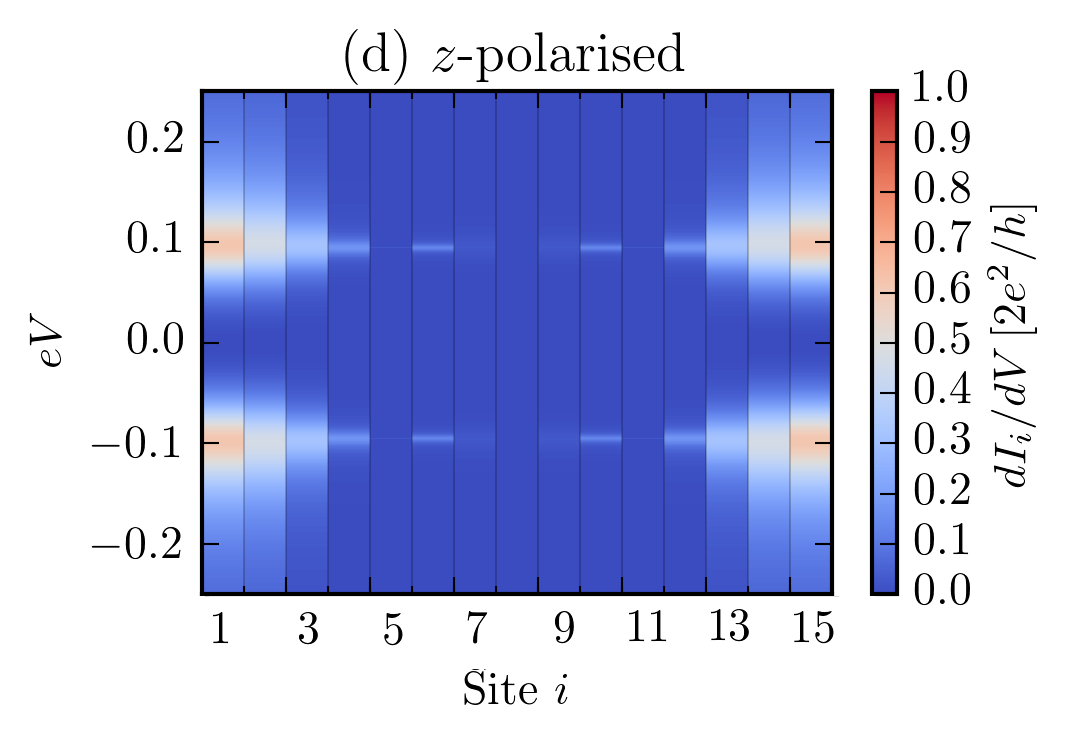}
\caption{Spatial tunneling conductance profile in the case of a short chain consisting of $N=15$ sites. (a) Profile at site $i=1$ for spin unpolarised ($\circ$) and
fully polarised tip along the $x,y,z$ axes. For a spin unpolarised tip one finds residual spectral weight for $V=0$. Due to the spin anisotropy in the MF induced
spin-polarisation we observe a significant difference in the height of the FBPs. When the polarisation is confined in the $xy$ plane or the tip is unpolarised, the
conductance is practically equal to $2e^2/h$, while for a tip magnetised along the $z$ axis one obtain a reduced height. We present the complete spatial profile for a 
fully spin-polarised tip along the $x,y,z$ axis in (b-d). Note the spectral weight asymmetry in b). Parameters: $\xi_0=80$, $k_F=6.0$, $\alpha=0.01$ and $M=0.85$.}
\label{fig:dIdVPanel2SingleMFshort}
\end{figure}

\noi as shown in Fig.~\ref{fig:dIdVPanel2SingleMFshort}. The tunneling conductance for these voltages reads
\bea
\left.\frac{dI_i}{dV}\right|_{\rm FBPs}=\frac{2e^2}{h}
\frac{\left(\Gamma_{aa}^i+\Gamma_{ee}^i\right)^2-4\Im^2\Gamma_{ae}^i}{\left(\Gamma_{aa}^i+\Gamma_{ee}^i\right)^2}\,.
\eea

\noi Notably the FBPs have a height equal to $2e^2/h$ only if $\Im\Gamma_{ae}^i=0$. Otherwise, the height is lower. As inferred by Fig.~\ref{fig:dIdVPanel1SingleMFshort}, 
{\color{black}we obtain an almost quantised conductance for all the cases except }when probing with a magnetic tip with polarisation along the $z$ axis. In this case, the
tunneling conductance is \textit{much weaker}. As a matter of fact, this is exactly the configuration employed in the experiment of Ref.~\onlinecite{Yazdani_Science}
and {\color{black} if our YSR model is applicable, our findings can provide one} possible explanation to the highly reduced signal, aside from the unavoidable temperature
broadening. Accor\-ding to our theory, orienting the spin-polarisation of the tip along the $x,y$ axes will drastically increase the conductance value to almost a single
quantum, while we additionally obtain that the ZBP can be found even for very short chains of $N=15$ sites. These remarkable findings are depicted in more detail in
Fig.~\eqref{fig:dIdVPanel2SingleMFshort} and demonstrate once again that SPSTM is a powerful tool for detecting these MF fingerprints attributed to the anisotropic MF
spin-polarisation.

Finally, if the coupling of the tip to $\gamma_e$ is completely negligible, then the tunneling conductance formula reads (see also Ref.~\onlinecite{FlensbergTunneling})
\bea
\frac{dI_i}{dV}=\frac{2e^2}{h}\frac{(eV)^2\left(\Gamma_{aa}^i\right)^2}{\left[(eV)^2-\delta\epsilon^2\right]^2+(eV)^2\left(\Gamma_{aa}^i\right)^2}\,.\label{eq:splitting}
\eea

\noi From the above expression we see that the ZBP \textit{disappears} and splits into two FBPs appearing for $eV=\pm \delta\epsilon$.

\subsection{One TSC chain with 2 MFs per edge}\label{Sec:dIdV2d02}

In this paragraph we examine the case of a single TSC magnetic chain where due to the preservation of chiral symmetry 2 MFs appear per edge as in Fig.~\ref{fig:dIdV2d02}. 
{\color{black}Although phases with two MFs per edge have not been experimentally demonstrated yet, they appear prominent to be realised in the near future. In fact, one can
engineer a 2MF per edge phase starting from a 1MF per edge phase, similar to the one discovered in Ref.~\onlinecite{Yazdani_Science}. As prescribed in
Ref.~\onlinecite{Interplay}, the latter can be achieved via a topological phase transition effected by varying a set of paramaters such as the adatom spacing, the SOC strength
or the applied magnetic field. 

\begin{figure}[b]
\centering
\includegraphics[scale=0.22]{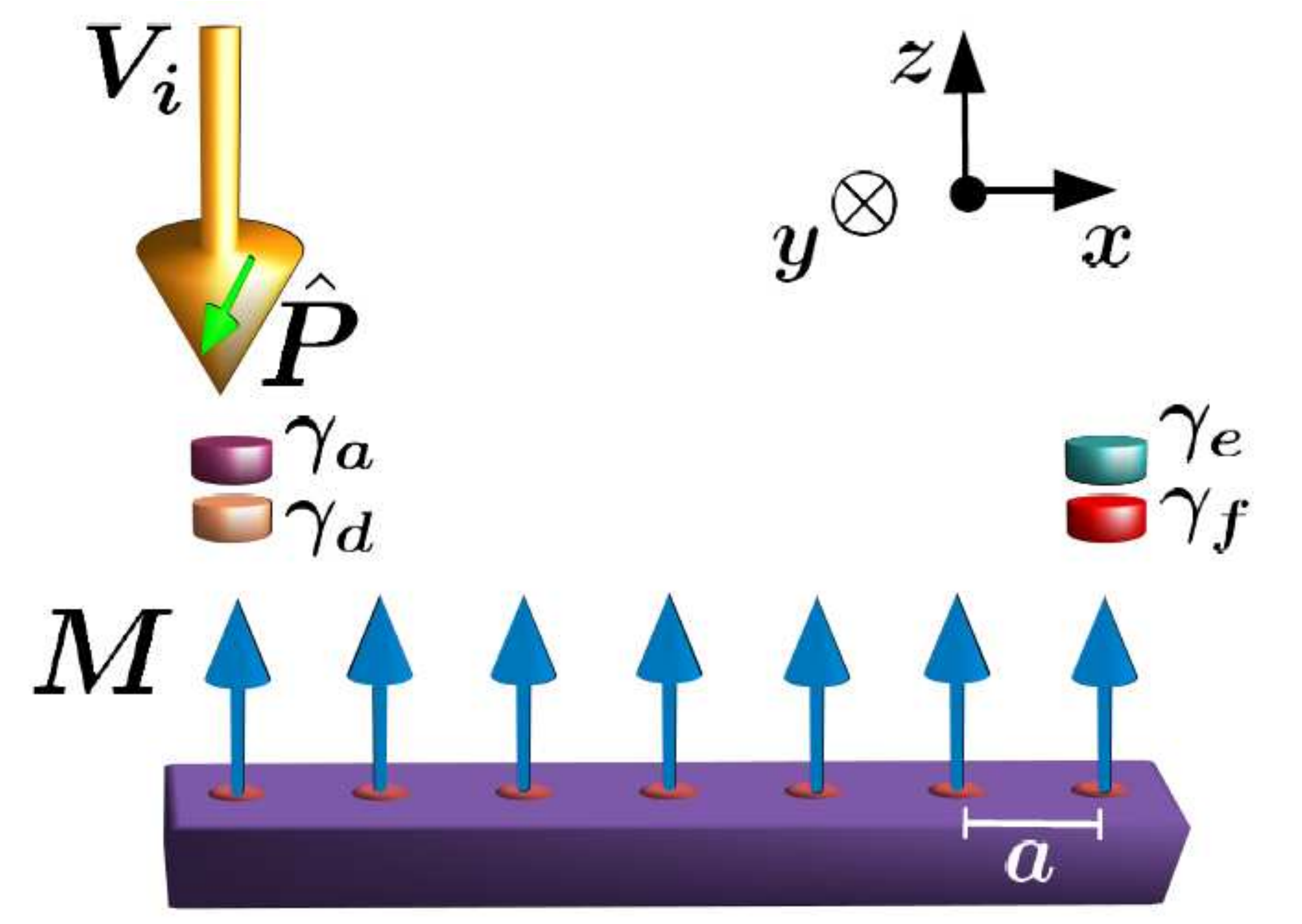}
\vspace{0.1in}
\caption{TSC harboring 2 MFs per edge protected by chiral symmetry. For sufficiently long chains solely the two MFs below the SPSTM probe tip need to be taken into account for
inferring the tunneling spectra. When owing a polarisation component along the $y$ axis, the magnetic tip itself can break locally chiral symmetry and hybridise the two MFs,
even in the absence of other symmetry breaking fields.}
\label{fig:dIdV2d02}
\end{figure}

Here we focus on a TSC magnetic chain in the 2MF phase with the two edges being infinitely separated, allowing us to restrict ourselves to the MF subspace of $\gamma_a$
and $\gamma_d$.} Essentially we exclude coupling between MFs of different edges and at the same time assume negligible coupling between the tip electrons and the MFs located
at the right edge. Consequently, both MFs on the left edge couple to the tip, while they can also couple to each other with a matrix element $m$, arising due to weakly broken
chiral symmetry. As we have already underlined the latter violation can be a consequence of the SPSTM tip itself, if the polarisation contains a component along the $y$ axis.
Under these conditions we have
\bea
\widehat{{\cal M}}=\left(\begin{array}{cc}0&m\\-m&0\\\end{array}\right)\quad{\rm and}\quad
\widehat{\Gamma}^i=\left(\begin{array}{cc}\Gamma_{aa}^i&\Gamma_{ad}^i\\(\Gamma_{ad}^i)^*&\Gamma_{dd}^i\\\end{array}\right)\,,\quad
\eea

\noi which are identical to the ones found in Eq.~\eqref{eq:MG01} with the correspondence $m\rightarrow\delta\epsilon$ and $d\rightarrow e$. Note that here the
off-diagonal elements for the linewidth functions are crucial and cannot be neglected a priori, since the two chiral symmetry protected MFs have spectral weight at the
same region. This is in stark contrast to Sec.~\ref{Sec:dIdV2d01} as also previous studies \cite{FlensbergTunneling} where the overlap of the 2 MFs involved can be
completely neglected in the infinite separation limit. The tunneling conductance can be obtained from Eq.~\eqref{eq:dIdV01} after performing the replacement
$\delta\epsilon\rightarrow m$ and $e\rightarrow d$. For $V=0$ one obtains
\bea
\left.\frac{dI_i}{dV}\right|_{V=0}=2\cdot\frac{2e^2}{h}\frac{\Gamma_{aa}^i\Gamma_{dd}^i-|\Gamma_{ad}^i|^2}
{\Gamma_{aa}^i\Gamma_{dd}^i-\Re^2\Gamma_{ad}^i+m^2}\,.
\eea

\begin{figure}[t]
\centering
\includegraphics[width=1\columnwidth]{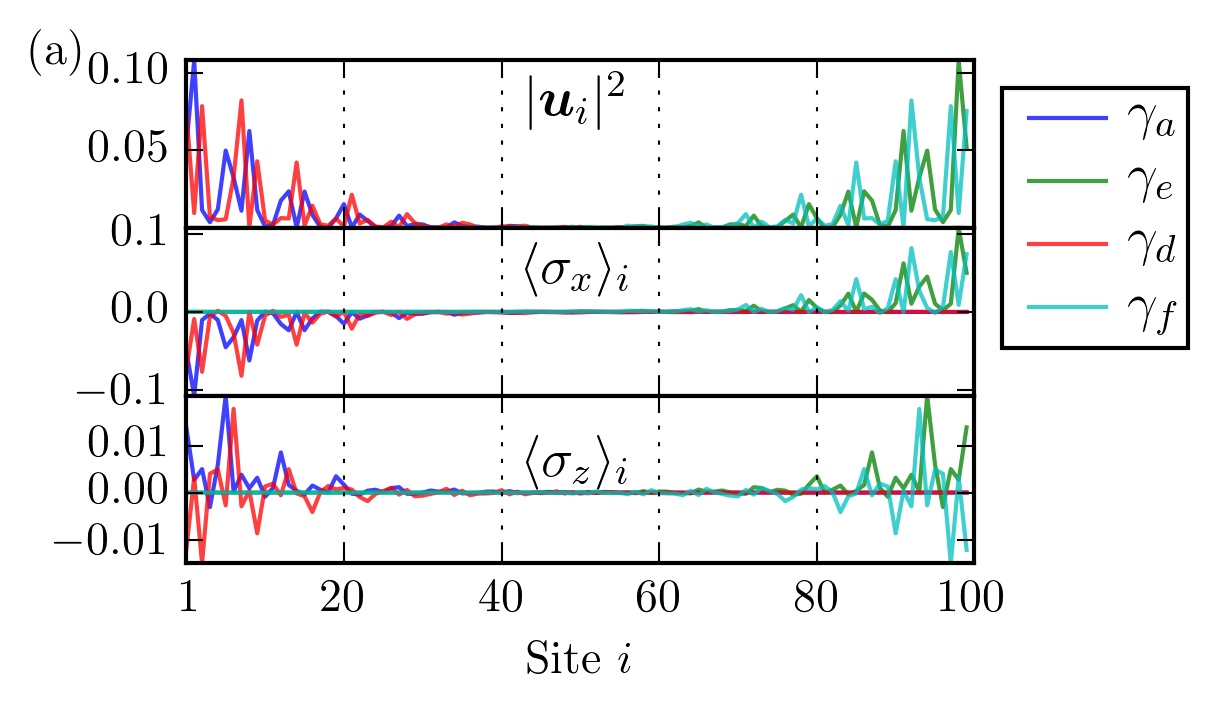}
\vspace{0.15in}
\includegraphics[width=0.5\columnwidth]{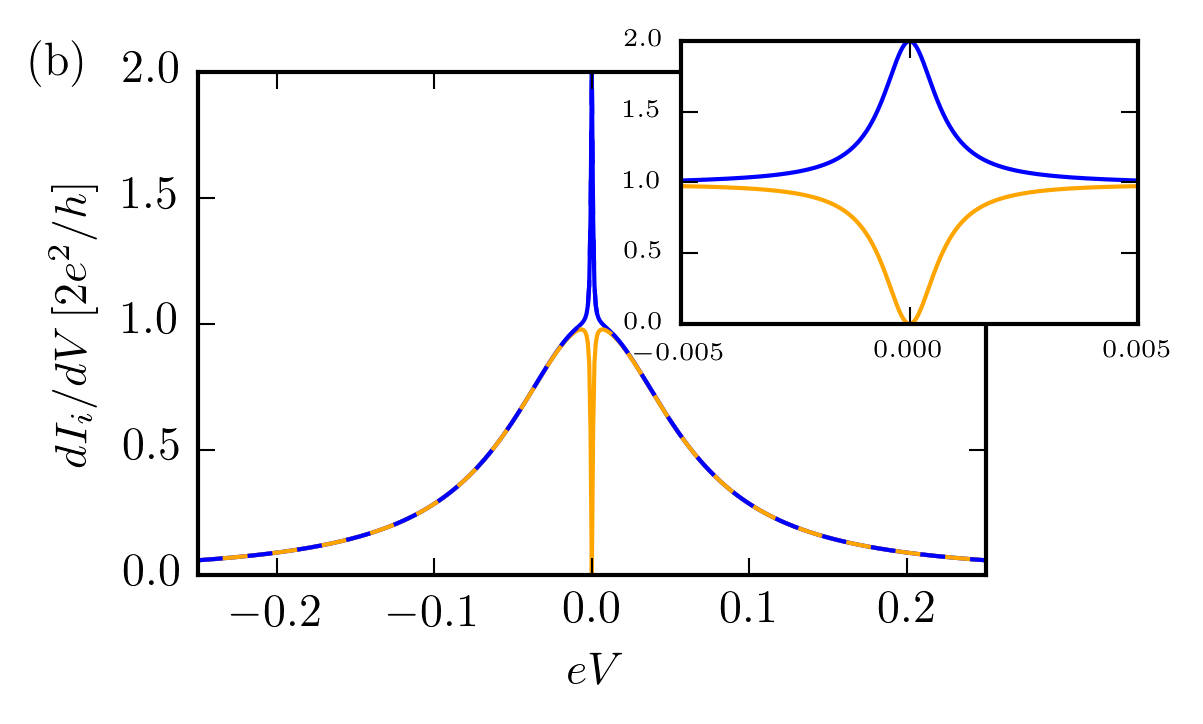}
\includegraphics[width=0.47\columnwidth]{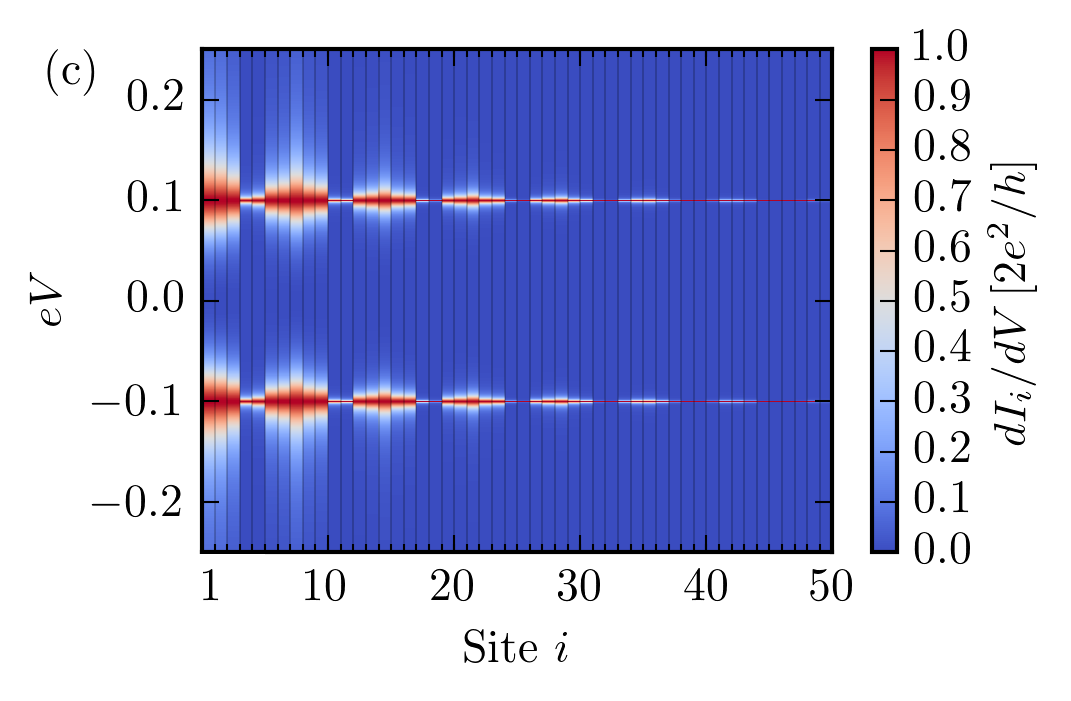}
\caption{(a) MF spectral weight $|\bm{u}_i|^2$ and electronic spin-polarisation. The spin-polarisation has only $x,z$ components. (b) Profile of the double unit of conductance
ZBP at $i=1$ when chiral symmetry is preserved. There exists a crossover voltage at which the spike-like profile switches to a broad hump-feature. This reflects the
contribution of two different sources to the conductance. The blue curve shows the full conductance, whereas the orange curve shows only the contribution of the second summand
in Eq.~\eqref{eq:summands}. In the inset we zoom around $V=0$. (c) Spatial tunneling conductance profile for broken chiral symmetry ($m=0.1$) and spin unpolarised tip.
Parameters: $N=100$, $\xi_0=80$, $k_F=6.0$, $\alpha=0.01$ and $M=0.85$.}
\label{fig:dIdVPanelDouble2MF}
\end{figure}

As previously, the ZBP still persists and here its height can be controlled in general by the chiral symmetry brea\-king field $m$ as also $\Im\Gamma_{ad}^i$. As shown in
Fig.~\ref{fig:dIdVPanelDouble2MF}a-b), the electronic spin polarisation of the MF wavefunctions is also in this case confined to the $xz$ plane. \textit{In addition, both
wavefunctions are real or imaginary.} Therefore, by employing a tip with polarisation along the $y$ axis one simultaneously induces finite values for $m$ and
$\Im\Gamma_{ad}^i$. For a spin-polarisation of the tip in the $xz$ plane or an unpolarised tip, $m=\Im\Gamma_{ad}^i=0$. Therefore, in the latter cases we obtain a ZBP with
double unit of conductance \cite{FlensbergTunneling}, as if the 2 MFs were unpaired \cite{KitaevUnpaired}.

When chiral symmetry is preserved ($m=\Im\Gamma_{ad}^i=0$), one obtains the following profile for a general voltage bias
\bea
&&\frac{dI_i}{dV}=\frac{2e^2}{h}\cdot\\
&&\frac{2\det^2(\widehat{\Gamma}^i)+(eV)^2\left[\left(\Gamma_{aa}^i\right)^2+\left(\Gamma_{ee}^i\right)^2+2(\Gamma_{ae}^i)^2\right]}
{\left[(eV)^2-\det \widehat{\Gamma}^i\right]^2+(eV)^2\left(\Gamma_{aa}^i+\Gamma_{ee}^i\right)^2}\,,\no\label{eq:summands}
\eea

One observes that only the first summand is responsible for the double unit of conductance ZBP, while the second term of the above equation contributes beyond a crossover
voltage where the sharp spike profile switches to a broad hump feature. Essentially, for very small vol\-tages the two MFs behave as unpaired and beyond the crossover voltage
they couple giving rise to two FBPs. This is obvious from the contribution of the second summand depicted with the orange line in Fig.~\ref{fig:dIdVPanelDouble2MF}b).

Note finally that there also special cases in which the presence of 2 MFs per edge can be even \textit{masked} and misinterpreted as a single MF per edge. In fact, if chiral
symmetry is preserved and for some particular values the condition $\Gamma_{aa}^i\Gamma_{dd}^i=(\Gamma_{ad}^i)^2=\Re^2\Gamma_{ad}^i$ additionally holds, then we obtain
the expression for the tunneling conductance
\bea
\frac{dI_i}{dV}=\frac{2e^2}{h}\frac{(\Gamma_{aa}^i+\Gamma_{dd}^i)^2}{(eV)^2+(\Gamma_{aa}^i+\Gamma_{dd}^i)^2}\,,
\eea

\noi which is identical to the one for a single MF per edge but with an effective broadening $\Gamma_{aa}^i+\Gamma_{dd}^i$. The special condition satisfied above implies that
essentially only one MF of the chiral symmetry protected MF pair is seen by the SPSTM tip.

\subsection{Two TSC chains with 1 MF per edge}\label{Sec:dIdV2d11}

\begin{figure}[b]
\centering
\includegraphics[width=1\columnwidth]{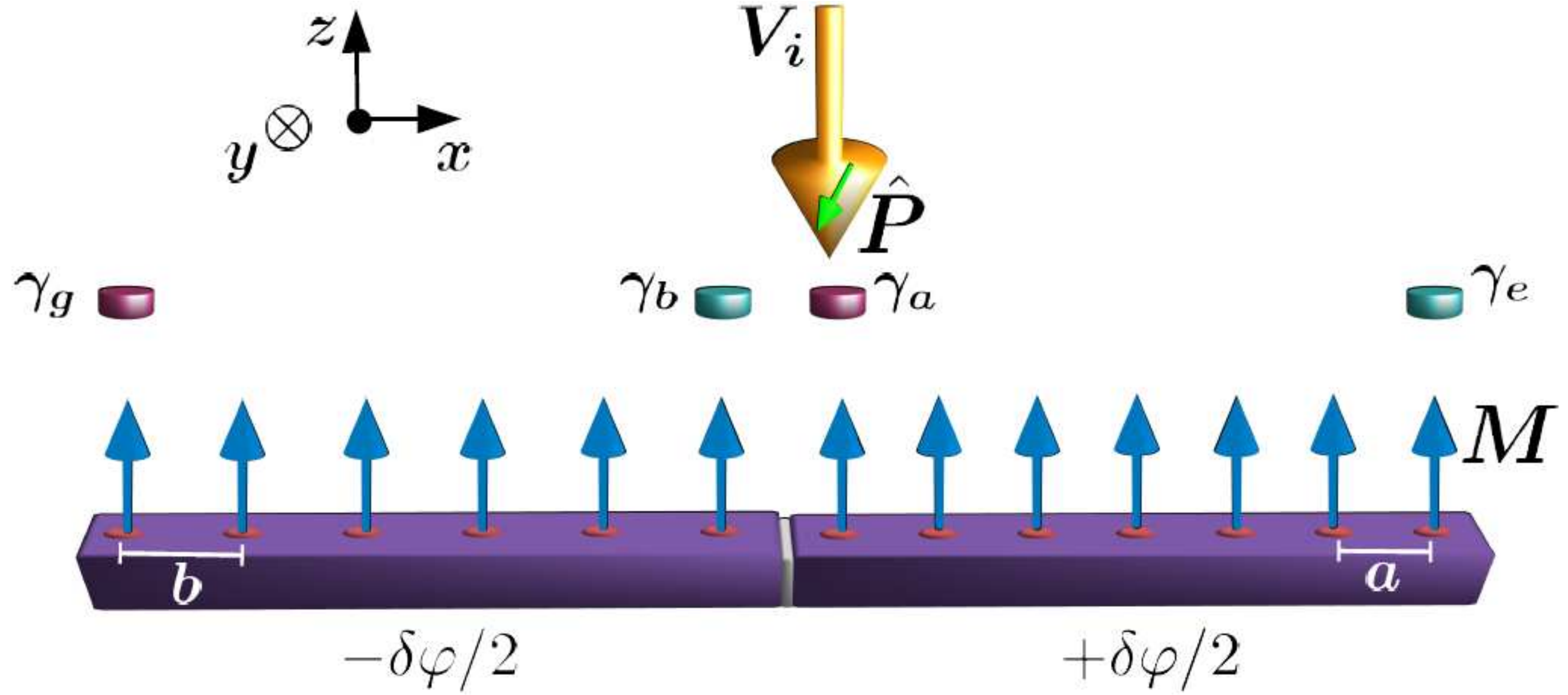}\\
\caption{Josephson junction of two tunnel-coupled TSC chains each of which harbors 1 MF per edge. For sufficiently long chains only the two MFs near the junction need
to be taken into account. Here only $\gamma_a$ couples to the SPSTM tip. The coupling $M$ of $\gamma_a$ and $\gamma_b$ has form of $4\pi$-periodic Josephson term.}
\label{fig:dIdV2d11}
\end{figure}

We now consider two coupled chains, each of which can harbor a single MF per edge, a situation depicted in Fig.~\ref{fig:dIdV2d11}. The particular setup can be useful for
indirectly probing the 4$\pi$-periodic Josephson effect. By assuming that the two chains are sufficiently long so that the MFs away from the junction can be excluded from
our analysis, we obtain
\bea
\widehat{{\cal M}}=\left(\begin{array}{cc}0&M\\-M&0\\\end{array}\right)\quad{\rm and}\quad
\widehat{\Gamma}^i=\left(\begin{array}{cc}\Gamma_{aa}^i&0\\0&0\\\end{array}\right)\,.\quad
\eea

\begin{figure}[t]
\centering
\includegraphics[width=0.49\columnwidth]{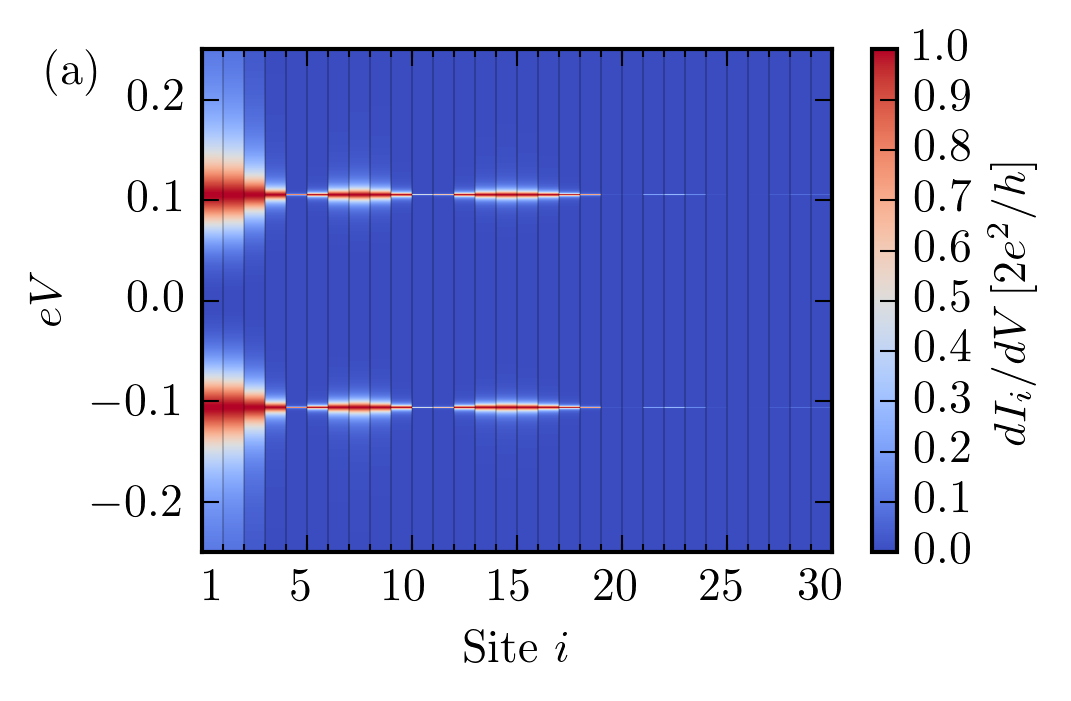}
\includegraphics[width=0.49\columnwidth]{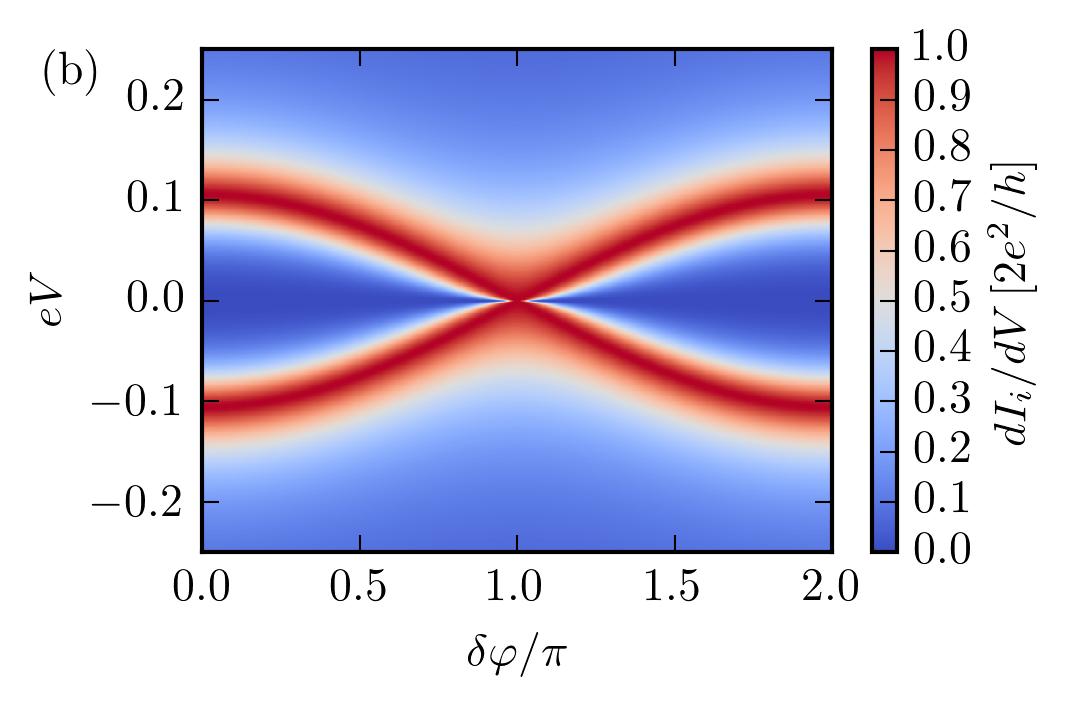}
\caption{(a) Spatial profile of the tunneling conductance for $\delta\varphi=0$. One obtain two FBPs at $eV=\pm M(\delta\varphi)$ (b) $2\pi$-periodic dependence of the
tunneling conductance ($i=1$) on the superconducting phase difference $\delta\varphi$. For $\delta\varphi=\pi$ one recovers the ZBP. We used a tunneling constant $t=0.1$ and
a decay length $l=10$ {\color{black} in units of $a=b=1$}.}
\label{fig:dIdVPanel1C1C1}
\end{figure}

\noi In this case the MF coupling matrix element originates from interchain tunneling and has the form
\bea
M=4\Im \sum_{i,j,\sigma} u^*_{i,\sigma,a}T_{i,j}e^{\imath(\varphi_i-\varphi_j)/2}u_{j,\sigma,b}\,.
\eea

Similarly to Eq.~\eqref{eq:splitting} one obtains for the particular case
\bea
\frac{dI_i}{dV}=
\frac{2e^2}{h}\frac{(eV)^2\left(\Gamma_{aa}^i\right)^2}{\left[(eV)^2-M^2\right]^2+(eV)^2\left(\Gamma_{aa}^i\right)^2}\,.
\eea

\noi Therefore, by imposing a difference $\delta\varphi$ between the phases of the two SCs, we can tune $M$ and modify the location of the FBPs. The tunneling conductance is a
$2\pi$-periodic function of $\delta\varphi$, resulting from the $4\pi$-periodic Josephson coupling $\propto\cos(\delta\varphi/2)$ between the $\gamma_{a}$ and $\gamma_b$.

\subsection{Two coupled TSC chains: one with 1 MF per edge (below the tip) and one with 2 MFs per edge}\label{Sec:dIdV2d21}

As previously we concentrate on the MFs near the junction. In this case only $\gamma_a$ couples to the tip and we have
\bea
\widehat{{\cal M}}=\left(\begin{array}{ccc}0&M_{ab}&M_{ac}\\-M_{ab}&0&m_{bc}\\-M_{ac}&-m_{bc}&0\end{array}\right)\,,
\widehat{\Gamma}^i=\left(\begin{array}{ccc}\Gamma_{aa}^i&0&0\\0&0&0\\0&0&0\end{array}\right)\,.\qquad
\eea

\noi The matrix elements $M_{ab}$ and $M_{ac}$ originate from interchain tunneling while $m_{bc}$ originates from chiral symmetry breaking only in the left or even in both
chains. Weak violation of chiral symmetry does not introduce any modi\-fi\-cation to the wavefunction of the $\gamma_a$ MF. Thus in the particular case, we obtain the
conductance formula
\begin{align}
\frac{dI_i}{dV}=\frac{2e^2}{h}
\frac{\left[(eV)^2-m_{bc}^2\right]^2\left(\Gamma_{aa}^i\right)^2}
{(eV)^2\left[(eV)^2-{\cal M}^2\right]^2+(\Gamma_{aa}^i)^2\left[(eV)^2-m_{bc}^2\right]^2}\,,
\end{align}

\noi with ${\cal M}=\sqrt{M_{ab}^2+M_{ca}^2+m_{bc}^2}$. For $V=0$ we observe that we obtain the ZBP. Indeed, this confirms the rule derived in
Ref.~\onlinecite{FlensbergTunneling} according to which the tunneling conductance of an odd number of coupled MFs demonstrate the ZBP. Moreover there are also two FBPs of
$2e^2/h$ at $eV=\pm {\cal M}$. However, if chiral symmetry is preserved, i.e. $m_{bc}=0$, we obtain
\bea
\frac{dI_i}{dV}=\frac{2e^2}{h}
\frac{(eV)^2\left(\Gamma_{aa}^i\right)^2}{\left[(eV)^2-M^2\right]^2+(eV)^2(\Gamma_{aa}^i)^2}\,.
\eea

\noi Essentially if chiral symmetry is preserved the system behaves as only 2 MFs become coupled, with an effective coupling $M=\sqrt{M_{ab}^2+M_{ac}^2}$. This becomes
tran\-spa\-rent by writing $\imath\gamma_a(M_{ab}\gamma_b+M_{ac}\gamma_c)=\imath M\gamma_a(M_{ab}\gamma_b/M+M_{ac}\gamma_c/M)$. The orthogonal linear combination of MFs
$M_{ac}\gamma_b-M_{ab}\gamma_c$ remains unpaired \cite{KitaevUnpaired} and unseen by the SPSTM tip. Therefore switching on and off the chiral symmetry breaking field can
controllably make the ZBP appear or disappear providing a smoking gun signature of MFs in these chains. On the other hand, one by controlling $\delta\varphi$ can shift the
two split peaks at $eV=\pm M$, adding another experimental knob for detecting MFs.

\begin{figure}[t]
\centering
\includegraphics[width=1\columnwidth]{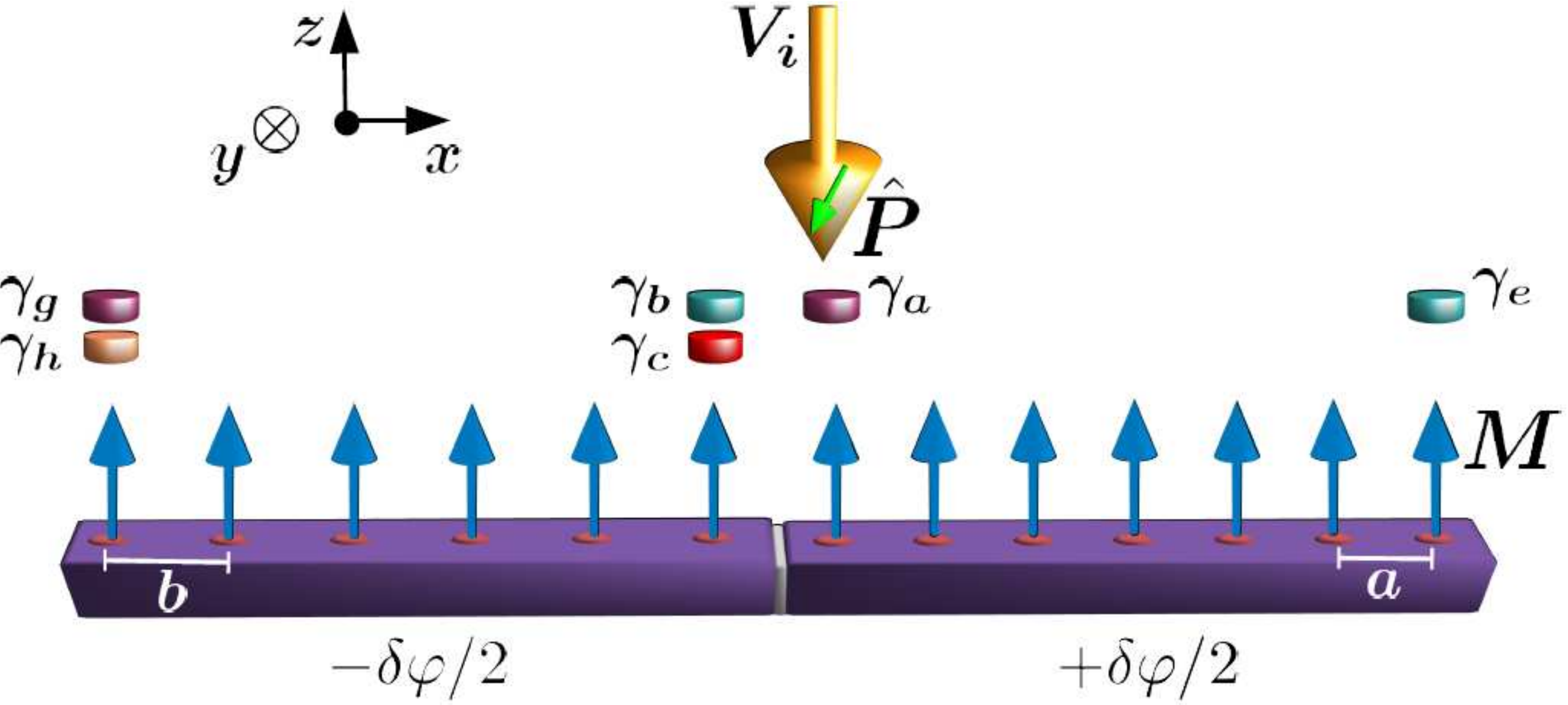}
\vspace{0.1in}
\caption{Josephson junction of two tunnel-coupled TSC chains harboring a different MF number per edge (two and one respectively). For sufficiently long chains, only
the three MFs near the junction need to be taken into account. Here only $\gamma_a$ couples to the SPSTM tip. The couplings $M$ of $\gamma_a$ to $\gamma_b$ and $\gamma_c$
have the form of $4\pi$-periodic Josephson terms. Chiral symmetry breaking can further mix $\gamma_b$ and $\gamma_c$.}
\label{fig:dIdV21}
\end{figure}

\begin{figure}[t]
\centering
\includegraphics[width=1\columnwidth]{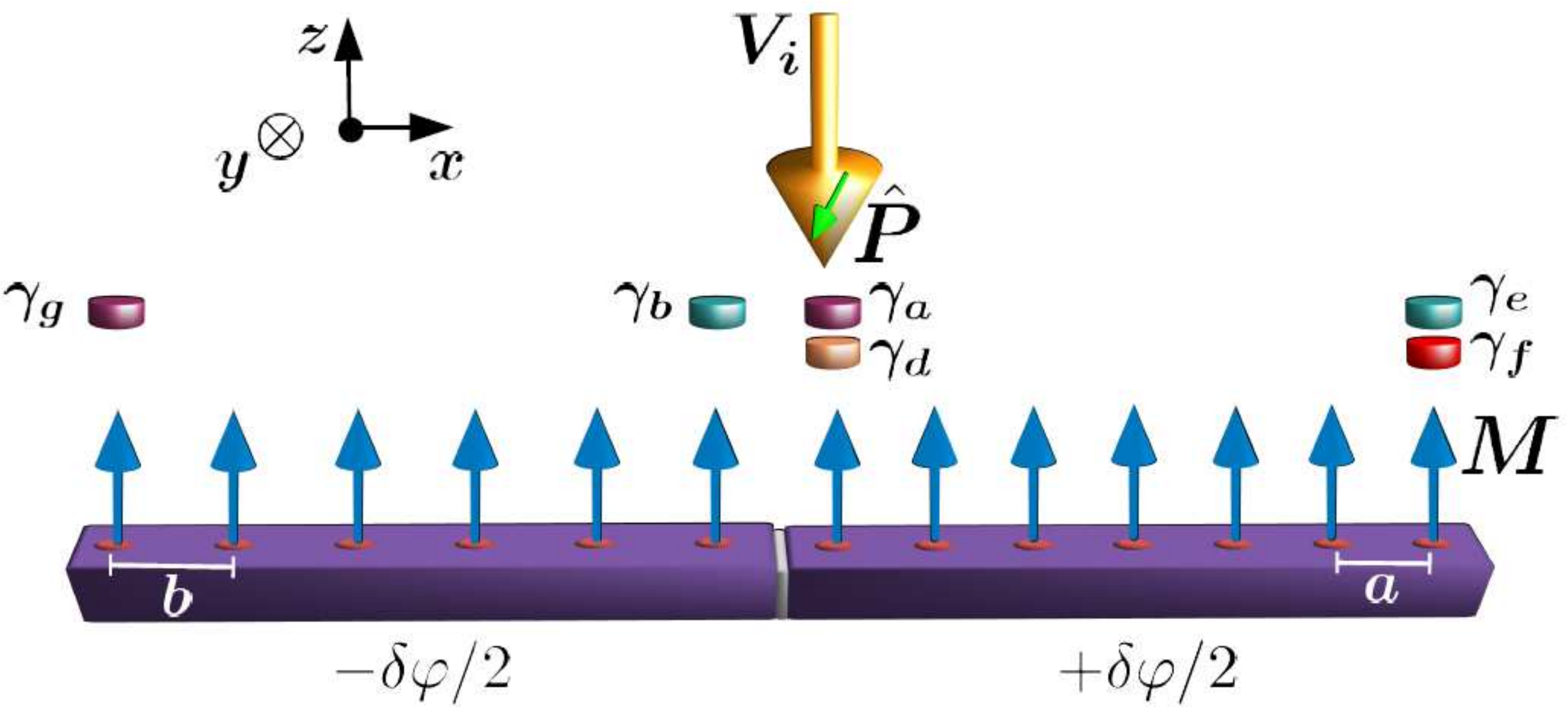}
\vspace{0.1in}
\caption{Josephson junction of two tunnel-coupled TSC chains harboring a different MF number per edge (one and two respectively). For sufficiently long chains, only
the three MFs near the junction need to be taken into account. Here only the chiral symmetry protected MF pair $\gamma_a$ and $\gamma_d$ couple to the SPSTM tip. The
couplings $M$ of $\gamma_b$ to $\gamma_a$ and $\gamma_d$ have the form of $4\pi$-periodic Josephson terms. If chiral symmetry is broken $\gamma_a$ and $\gamma_d$ can also
have a finite coupling.}
\label{fig:dIdV2d12}
\end{figure}

\subsection{Two coupled TSC chains: one with 1 MF per edge and one with 2 MFs per edge (below the tip)}\label{Sec:dIdV2d12}

In the present paragraph we consider the second possible configuration for two chains with uneven edge MF number, as shown in the Fig.~\ref{fig:dIdV2d12}. In the
particular case one obtains the matrices
\begin{align}
\widehat{{\cal M}}=\left(\begin{array}{ccc}0&m_{ad}&M_{ab}\\-m_{ad}&0&M_{db}\\-M_{ab}&-M_{db}&0\end{array}\right)\,,
\widehat{\Gamma}^i=\left(\begin{array}{ccc}\Gamma_{aa}^i&\Gamma_{ad}^i&0\\(\Gamma_{ad}^i)^*&\Gamma_{dd}^i&0\\0&0&0\end{array}\right)\,.\qquad
\end{align}

\noi The matrix elements $M_{ab}$ and $M_{db}$ originate from interchain tunneling while $m_{ad}$ originates from chiral symmetry breaking in the particular or even both
chains. Weakly breaking chiral symmetry does not affect the wavefunction of the single edge MF.

Here the resulting tunneling conductance expression is lengthy and therefore we present it in the Appendix. However, one can directly infer that the ZBP persists \textit{both}
in the presence or absence of chiral symmetry. The reason is that the MFs below the tip become always indirectly coupled via their interaction with the tip, and thus the
presence or not of chiral symmetry is only important for ensuring the presence of the 2 MFs. Since all three MFs are coupled and their number is odd the appearance of
the ZBP was expected according to the rule of Ref.~\onlinecite{FlensbergTunneling}. If the couplings to $\gamma_b$ become zero, then we return to the case of 2 MFs at the edge
of a single chain discussed in Sec.~\ref{Sec:dIdV2d02}, in which case the existence of a ZBP crucially depends on the persistence of chiral symmetry.

\subsection{Two TSC chains: Both with 2 MFs per edge}\label{Sec:dIdV2d22}

For completing our analysis we proceed with examining the case in which both chains harbor 2 MFs per edge and are tunnel-coupled. By focusing only on the MFs near the
junction depicted in Fig.~\ref{fig:dIdV3d} we can write
\bea
\widehat{{\cal M}}=\left(\begin{array}{cccc}0&m_{ad}&M_{ac}&M_{ab}\\-m_{ad}&0&M_{dc}&M_{db}\\-M_{ac}&-M_{dc}&0&m_{cb}\\-M_{ab}&-M_{db}&-m_{cb}&0\end{array}\right)\,,
\qquad\qquad\\\no\\
\widehat{\Gamma}^i=\left(\begin{array}{cccc}\Gamma_{aa}^i&\Gamma_{ad}^i&0&0\\(\Gamma_{ad}^i)^*&\Gamma_{dd}^i&0&0\\0&0&0&0\\0&0&0&0\end{array}\right)\,,
\widehat{M}=\left(\begin{array}{cc}M_{ac}&M_{ab}\\M_{dc}&M_{db}\end{array}\right)\,.\quad
\eea

\noi The expression that one obtains for the tunneling conductance is lengthy and therefore we do not include it in this manuscript. However it is easy to retain the
expression for the tunneling conductance at $V=0$, which is sufficient for providing information about the qualitative characteristics of the system. The zero voltage
conductance reads
\begin{align}
\left.\frac{dI_i}{dV}\right|_{V=0}=\frac{2e^2}{h}\frac{2m_{cb}^2\det\widehat{\Gamma}^i}
{\left(\det\widehat{M}-m_{ad}m_{cb}\right)^2+m_{cb}^2\det(\Re\widehat{\Gamma}^i)}\,.
\end{align}

It is straightforward to observe that we obtain a ZBP but the conductance does not take a quantised value. At this point we can further focus on special cases. If chiral
symmetry is at least restored for the chain which does not couple to the SPSTM tip (i.e. $m_{cb}=0$), the ZBP va\-ni\-shes, implying that effectively an even number of MFs out
of the four are essentially probed by the tip. On the other hand, if chiral symmetry is only restored for the chain probed by the tip (i.e. $m_{ad}=0$ and $P_y=0$), the ZBP
remains but with modified and still non-quantised conductance value. Note however that if $m_{ad}=0$ and $\det\widehat{M}=0$ then we obtain a peak of a double unit of
conductance as if only two unpaired MFs appear in the system. Finally, if only $\det\widehat{M}=0$, the tunnel coupling matrix has a zero eigenvalue implying that the
tip does not see the second chain and we return to the case of 2 MFs probed simultaneously by the tip.

\section{Tunneling conductance beyond the MF-induced Andreev processes}\label{Sec:beyond}
{\color{black}
Throughout the whole manuscript the tunneling conductance was computed based on the approximation that the YSR operators can be replaced with the MF operators, i.e.,
$\psi_{i,\sigma}=\sum_nu_{i,\sigma,n}\gamma_n$. Essentially, we kept only MF-induced Andreev processes. In fact, these are the only possible Andreev processes which can occur,
since the SPSTM is considered non-superconducting in the present discussion (see e.g. \onlinecite{AndreevProcesses}). The remaining contribution to the conductance can only
arise by single electron processes. Due to the bulk gap in the energy spectrum of the TSC chain, single electron tunneling is suppressed and can only occur if some YSR states
become unoccupied, so that the tip-electrons can tunnel into them. This can happen for finite temperatures or/and due to inelastic scatering with collective modes in the
substrate such as phonons. For more details see Ref.~\onlinecite{SCtip}, where it has been also shown that the inelastic processes are thermally activated. Therefore the
applicability of our results is restricted only to very low temperatures where the Andreev approximation is well justified, while in addition it is required for the SPSTM tip
to be located very close to the substrate.}

\section{Conclusions and perspectives}\label{Sec:Conclusions}

We explored new distinctive MF features which can be measured via spin-polarised scanning tunneling microscopy in hybrid devices consisting of ferromagnetic chains on top of
spin-orbit coupled superconductors. For the calculations we adopted a microscopic model de\-scri\-bing Yu-Shiba-Rusinov chains which can harbor 1 or 2 MFs per edge if chiral
symmetry is present. 

For an isolated topological YSR chain with a single MF per edge, we showed that the tunneling conductance delicately depends on the direction along which the tip is
spin-polarised. If fact, for a fully spin-polarised tip there can be special angles of the tip polarisation for which the conductance vanishes. In addition, in the
case of short chains where MFs on both edges contribute, one finds that depending on the tip polarisation the signal can be extremely weak while at the same time other
directions support an almost quantised ZBP. This can be relevant for the recent measurements of Ref.~\onlinecite{Yazdani_Science} which were charac\-te\-ri\-sed by a weak
signal. According to our findings, the polarisation and Zeeman field of the ex\-pe\-ri\-ment (i.e. $z$ axis) were oriented along the direction for which the MF signal appears
to be the lowest. Within our scenario, altering the field angle could lead to signal enhancement and allow the possible observation of the long-sought-for ZBP. 

Moreover, we showed that in the case of a single chain with 2 symmetry protected MFs per edge, the tip (or an additional Zeeman field along the $y$ axis) can induce a weak
chiral symmetry violating term which can controllably modify the tunneling spectra. In addition, for the case of tunnel-coupled topological chains, one can induce a difference
between the phases of the two superconductors in order to modify the location of emer\-ging finite bias peaks via the $4\pi$-Josephson coupling. In fact, the tunneling
conductance could be used as an indirect probe of the latter. Furthermore, for two chains with diffe\-rent edge MF number, tunable chiral symmetry violation and restoration
can switch on and off the ZBP which is a robust MF feature. 

{\color{black} Note that in experiments based on self-assembled magnetic chains, as in Ref.~\onlinecite{Yazdani_Science}, junctions can be currently difficult to fabricate.
Nonetheless, a number of the above mentioned Josephson effects can be still accessed by either inducing a supercurrent flow along a short TSC magnetic chain (see e.g.
Refs.~\onlinecite{Heimes,Romito}) or employing a chain in a ring geometry through which flux can be threaded. In both configurations the MFs of the left and right edges are
coupled and feel different superconducting phases, thus behaving similarly to the neighbouring MFs of two tunnel-coupled chains.} 

Conclusively, this novel MF characteristics relying on the MF spin-polarisation, which were extracted from a realistic YSR microscopic model for these chains, reveal new
experimental methods for unambiguously detecting MFs in the near future. 

\begin{acknowledgments}

PK is grateful to A. Shnirman and S. Kourtis for fruitful and insightful discussions. 

\end{acknowledgments}

\newpage

\begin{widetext}

\appendix*

\section{Expression for the tunneling conductance for the case presented in Sec.~\ref{Sec:dIdV2d12}}

\bea
\frac{dI_i}{dV}=\frac{2e^2}{h}\frac{N(eV)}{D(eV)}\,,
\eea

\noi where we have introduced the denominator
\bea
D(\omega)&=&\omega^6+\omega^4\left[(\Gamma_{aa}^i)^2+(\Gamma_{dd}^i)^2+2\Re^2\Gamma_{ad}^i-2{\cal M}^2\right]\no\\
&+&\omega^2\left\{\left[\Gamma_{aa}^i\Gamma_{dd}^i-\Re^2\Gamma_{ad}^i+{\cal M}^2\right]^2
-2(\Gamma_{aa}^i+\Gamma_{dd}^i)\left[\Gamma_{aa}^i M_{db}^2+\Gamma_{dd}^i M_{ab}^2-2(\Re\Gamma_{ad}^i)M_{ab}M_{db}\right]\right\}\no\\
&+&[\Gamma_{aa}^i M_{db}^2+\Gamma_{dd}^i M_{ab}^2-2(\Re\Gamma_{ad}^i)M_{ab}M_{db}]^2\,,
\eea

\noi and the nominator
\bea
&&N(\omega)=\omega^4\left\{(\Gamma_{aa}^i)^2+(\Gamma_{dd}^i)^2+2\left[\Re^2\Gamma_{ad}^i-\Im^2\Gamma_{ad}^i\right]\right\}+
2\omega^2\left[\Gamma_{aa}^i\Gamma_{dd}^i-\Re^2\Gamma_{ad}^i+m_{ad}^2\right]\left[\Gamma_{aa}^i\Gamma_{dd}^i-|\Gamma_{ad}^i|^2\right]\no\\
&&-2\omega^2\left\{
\left[(\Gamma_{aa}^i)^2+\Re^2\Gamma_{ad}^i-\Im^2\Gamma_{ad}^i\right]M_{db}^2+
\left[(\Gamma_{dd}^i)^2+\Re^2\Gamma_{ad}^i-\Im^2\Gamma_{ad}^i\right]M_{ab}^2-
2(\Gamma_{aa}^i+\Gamma_{dd}^i)(\Re\Gamma_{ad}^i)M_{ab}M_{db}\right\}\no\\
&&+[\Gamma_{aa}^i M_{db}^2+\Gamma_{dd}^i M_{ab}^2-2(\Re\Gamma_{ad}^i)M_{ab}M_{db}]^2\,,
\eea

\noi with ${\cal M}=\sqrt{m_{ad}^2+M_{ab}^2+M_{db}^2}$.

\end{widetext}

\end{document}